\documentclass[%
 reprint,
 amsmath,amssymb,
 aps,
]{revtex4-1}

\usepackage{graphicx}% Include figure files
\usepackage{float}
\usepackage{dcolumn}% Align table columns on decimal point
\usepackage{bm}% bold math
\usepackage{color}
\usepackage{multirow}% http://ctan.org/pkg/hhline
\usepackage[normalem]{ulem} %for strike text
\usepackage{amsmath}
\usepackage{array}
\usepackage{booktabs}

\begin{document}
\newcommand*{\rom}[1]{\expandafter\@slowromancap\romannumeral #1@}

%\preprint{APS/123-QED}

%===========================================================================
\title{Controlling polymer translocation with crowded medium and polymer length asymmetry}

\author{Vrinda Garg$^{1}$}
\author{Rejoy Mathew$^{1}$}
\author{Riyan Ibrahim$^{1}$}
\author{Kulveer Singh$^{1}$}
\author{Surya K. Ghosh$^{1}$}
\email{skghosh@nitw.ac.in}

\affiliation{%
$^{1}$Department of Physics, National Institute of Technology, 
 Warangal, 506004, India}

\date{\today}
%\date{ February 23, 2024}

\begin{abstract}
Polymer translocation in crowded environments is a ubiquitous phenomenon in biological systems. We studied polymer translocation through a pore in free, one-sided (asymmetric), and two-sided (symmetric) crowded environments. Extensive Langevin dynamics simulation is employed to model the dynamics of the flexible polymer and crowding particles. We studied how crowding size and packing fraction play a crucial role in the translocation process. After determining the standard scaling properties of the translocation probability, time, and MSD, we observed that the translocation rate and bead velocities are location-dependent as we move along the polymer backbone, even in a crowd-free environment. Counter-intuitively, translocation rate and bead velocities showed the exact opposite behavior; for example, middle monomers near the pore exhibit maximum bead velocity and minimum translocation rate. Free energy calculation for asymmetrically placed polymer indicates there exists a critical number of segments that make the polymer to prefer the receiver side for translocation. For one-sided crowding, we have identified a critical crowding size above which there exists a non-zero probability to translocate into the crowding side instead of the free side. Moreover, we have observed that shifting the polymer towards the crowded side compensates for one-sided crowding, yielding an equal probability akin to a crowder-free system. Under two-sided crowding, the mechanism of how a slight variation in crowder size and packing fraction can force a polymer to switch its translocation direction is proposed, which has not been explored before. Using this control we achieved an equal translocation probability like a crowd-free scenario. These conspicuous yet counter-intuitive phenomena are rationalized by simple theoretical arguments based on osmotic pressure and radial entropic forces.

%====================================================================

%\begin{description}
%\item[PACS numbers] 87.15.A-, 87.10.Vg, 87.10.Rt, 87.10.Mn, 87.10.Hk
%\end{description}
\end{abstract}

%\pacs{} 

\maketitle

%=========================================================================
%=========================================================================

\section{Introduction}

The transport of a polymer through a pore is omnipresent in many biological systems. Examples include RNA passing through a pore created by a membrane-bound protein \cite{butler2006determination,dubbeldam2007polymer}, RNA and DNA sequencing \cite{meller2000rapid, howorka2001sequence, vercoutere2001rapid, SinghGhoshEPL2012, nakane2004nanosensor,muthukumar2001translocation}, polymer transport processes \cite{zhou2022molecular,citovsky1993transport,SinghEPL2013, rajarshiChakrabarti2022migration}, gene therapy \cite{di1997phase, doi1988theory, sung1996polymer}, viral ejection \cite{kasianowicz2002physics, peskin1993cellular, kasianowicz1996characterization, ChandranJBSD2023}, controlled delivery \cite{liechty2010polymers, bettini2001translocation, MaityJDDS2023}, and polymer sorting and ultrafiltration \cite{Wu2010MacroSorting}. In in-vitro setups, translocations are mainly induced by driving forces, such as external applied electric field, \cite{ali2005polymer, kasianowicz1996characterization, lubensky1999driven, sarabadani2022driven, Hsiao2020ACSO_Translocation, buyukdagli2019theoretical, lathrop2010monitoring}, controlling the translocation of a single molecule through a glass nanopore on a 3D nanopositioner \cite{leitao2023spatially, rajarshiChakrabarti2023structure}, pulling force exerted on polymer's end \cite{kantor2004anomalous, ollila2009polymer, panja2008passage}, binding particles (chaperones) \cite{ambjornsson2004chaperone, luo2006polymer, yu2020translocation, yu2011chaperone}, are known as forced translocation \cite{LUBENSKY19991824}. In many biological systems, unforced translocations occur naturally dictated by conformational entropy \cite{kasianowicz1996characterization, kasianowicz2002physics, lehtola2010unforced, SimonPNAS1992}. Despite its omnipresence and importance in biological processes, unforced translocation has received less attention \cite{lehtola2010unforced, SimonPNAS1992}. Especially the system's intrinsic parameters, such as the symmetry of the length distribution of the polymer being translocated and the effects of crowding present in the cellular environment. Here, we investigate the generic behavior of self-avoiding flexible polymers\cite{GhoshPRE2009} exhibiting unforced translocation through a pore for free and crowded environments. A deeper understanding of these natural unforced processes can help in designing more efficient practical in-vitro setups. 

Polymer translocation through pores displays a broad range of scaling regimes, as shown by numerical simulations \cite{luo2010polymerJCP,wong2008scaling,lucas2022unbiased,luo2008dynamical}, analytical theories \cite{gopinathan2007polymer,wen2021current}, and experiments \cite{kasianowicz1996characterization}. The average translocation time $\tau$ as a function of the chain length $N$ is an important measure of the underlying dynamics. Especially in the case of unforced translocation, the barrier is so high that it is almost practically impossible to have a successful translocation solely due to the thermal agitation for long polymer chains in the limit of $N \gg 1$ \cite{kantor2004anomalous}. As the repeated chemical units of the polymer pass through the pore, they encounter a depletion in their available conformational entropy creating an overall entropic barrier resulting in constrained diffusion. According to Kramers's analysis of diffusion across an entropic barrier, the translocation time is scaled as $\tau \sim N^{\alpha}$, for phantom chains in the case of unforced translocation the scaling exponent is $\alpha =2$ ($\tau \sim N^2$) and for the forced translocation $\alpha = 1$ ($\tau \sim N$) \cite{kramers1940brownian}. The Rouse model for the dynamics of phantom chains also predicts a time of the order of $N^2$ for equilibration. For a phantom chain, translocation time depends on the relative magnitudes of three-time scales in terms of dimensionless factor that characterizes the translocation process $\Tilde{\tau}$ and medium's viscosity $\eta$: total translocation time $\tau \sim (b^2/D_p)\Tilde{\tau}$, polymer relaxation time $\tau_R \sim (\eta b^3/k_BT)N^2$, and $\tau_0 \sim \phi^{-2/3}/D_0$ for obstacles motion, where Gaussian polymer (phantom chain) of length $N$ (in units of the Kuhn length $b$) has diffusion coefficient $D_p$, $D_0$ is the diffusion coefficient of obstacles, and $\phi$ is the volume fraction of randomly distributed obstacles. Two regimes exist depending on the relative mobility of the obstacles: the dynamic obstacles regime where $\tau_0 \ll \tau_R \ll \tau$ and the static obstacles regime where $\tau_R \ll \tau \ll \tau_0$ \cite{sung1996polymer, doi1988theory, gopinathan2007polymer, chen2013dynamics, kantor2004anomalous, lehtola2010unforced, luo2010polymerPRE}. With excluded volume, in the presence of an external driving force $f$, the translocation time $\tau$ scales as $\tau \sim N^{\beta} / f^{\gamma}$ where $\gamma$ is the scaling exponent associated with the external forces. Biased translocation exhibits anomalous behavior with various values of both these exponents in several experimental \cite{BrantonEP2022, molcrette2022experimental} and computational studies \cite{chuang2001anomalous, afrasiabian2020journey}. Once self-avoidance is included, translocation time increases dramatically with a scale of $\alpha =  2.5$ ($\tau \sim N^{2.5}$), which is an exponent of Rouse relaxation of self-avoiding chain in 2D \cite{kantor2004anomalous,chuang2001anomalous}. For the self-avoiding chain, the translocation time is longer than the Rouse relaxation time ($\tau >> \tau_R$). According to the flory-exponent theory, the scaling of translocation time for a self-avoiding chain is given as  $\tau$ $\sim$ $N^\beta$ where $\beta$ = 1 + 2 $\nu$ and with the flory exponent $\nu = 0.75$ ($\tau \sim N^{2.5}$) in 2D and $\nu = 0.588$ ($\tau \sim N^{2.176}$) in 3D \cite{luo2006polymer, luo2010polymerPRE, chen2013dynamics}. Likewise, the scaling for MSD has been noted. It takes the power law form: $\langle \Delta r^{2} (t)\rangle \sim t^{\alpha}$, where $\alpha$ is the anomalous diffusion exponent. It is observed that $\alpha$ = 0.8 resembles the experimental value \cite{lehtola2010unforced, chuang2001anomalous,chaudhury2008model}. The shapes and sizes of the fluidic channel and cavity 
\cite{polson2019polymer} also have a strong effect when a polymer is translocated into or out of a confined environment \cite{luo2010polymerJCP, luo2009polymerPRE, luo2010polymerPRE, sarabadani2022driven}. 
 
In this work, we considered a self-avoiding flexible polymer in an unbiased and unforced environment, threading its way through the pore as a fundamental model and as a first step towards understanding the translocation dynamics in-vivo set-ups. Here, we specifically focused on the system's intrinsic parameters rather than the external influences. We examined certain parameters to elucidate translocation dynamics in this system thoroughly. Quantitative measurements of the probability of translocation events and the corresponding translocation times through the pore provide a fundamental basis for our study and validate the process. To understand the diffusion mechanism, we investigate the MSD of the center of mass of polymer, head, middle, and tail monomers. A study on how different segments of the polymer are moving across the pore is drawn to attention. We examine the quantities, such as the average bead crossing time of individual monomers, the translocation rate of each bead, and their translocation speed across the pore. In particular, we find a counterintuitive trend of monomers' translocation rate and translocation speed as they pass through the pore.

As the polymer translocates through the pore, the number of accessible conformations significantly reduces. This leads to a decrease in the chain's conformational entropy and an increase in its free energy. We examined the free energy profile for the no-crowding case and observed a symmetric free energy barrier when the polymer chain is placed symmetrically at the pore. This analysis resembles the analytical calculations \cite{muthukumar2016polymer}. Further, we find that an asymmetrically placed polymer leads to a critical number of segments that, when translocated, causes the free energy minima and polymer to prefer the receiver side.

The highly crowded environment of actual biological cells containing large macromolecules like proteins, lipids, ribosomes, and cytoskeleton fibers features volume occupancies up to $\phi \sim 40 \%$. Studying the impact of crowding is crucial for comprehending translocation in realistic cellular environments due to high cell density\cite{muthukumar2016polymer, gopinathan2007polymer, polson2019polymer, shin2015polymer}.  A number of recent works have shed light on the translocation processes in the presence of non-inert crowder (chaperons)\cite{yu2011chaperone, ambjornsson2004chaperone}. An aspect that has received very little attention is the impact of inert crowding, resulting in a biased but still unforced system by introducing crowders at one side of the box, leading to asymmetric crowding and on both sides of the box, creating a symmetric crowded environment as a second step to understand the translocation dynamics in crowded in-vivo set-ups. Polymer threading its way through such an environment is subjected to an entropic penalty, and interaction between crowder-polymer affects the translocation dynamics. Here, we quantify the effects of modifying the length distribution of polymer from its initial configuration, varying the size and packing fraction of the crowders, and show that it significantly affects the translocation dynamics. We find that when the crowders size reaches a crossover and deviates from the size of each chemical unit of the polymer, the polymer-crowder interaction fundamentally changes the direction of translocation and shows a sudden jump within two extremes of probability. Moreover, we show that by tuning the crowded environment, we can control the dynamics and switch the direction of the translocation process, which can be applied for better drug delivery, DNA sequencing, and transport processes.

The article is constructed as follows: Section II elucidates the methods and models employed in this study. In section III, we present the simulation results and discussions for the three distinct modeling frameworks. Firstly (section III A), we examine the dynamics of polymer translocation without crowder.  Next (section III B), we investigate the impact of the crowder on one side of the box. Finally (section III C), we study the translocation process driven by the crowders on both sides of the box. Section IV summarizes the conclusions drawn from this work. 

\begin{figure*}[ht]
  \includegraphics[width=\textwidth,height=4cm]{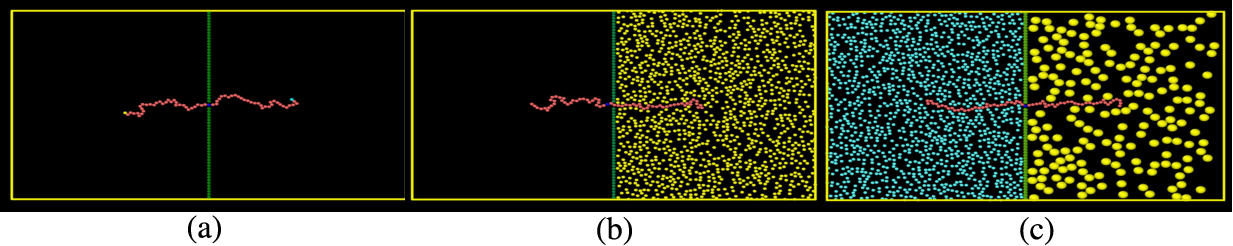}
  \caption{Schematic illustration of the system of polymer translocation process through a pore. The wall is purely repulsive and consists of immobile LJ particles, and has a width $\sigma$ the same as the size of each repeated unit of polymer (monomer). As an initial configuration, the polymer is placed symmetrically at the pore center. The left part represents the \textit{cis}-translocation of polymer, and the right part represents the \textit{trans}-translocation of polymer of length, $N = 65$. Here, we show the initial configuration of all the systems with a polymer being symmetrically placed at the pore. (a) Polymer Translocation in a crowd-free environment. (b) Polymer translocation in the presence of one-sided crowding. Yellow particles on the trans side represent crowders. (c) Polymer translocation when crowders are present on both sides of the box. Packing fractions of crowders on both sides is kept the same. We are changing the size of the crowders on the trans side, which is depicted in yellow, while the crowders on the cis side are shown in blue.  }
  \label{fig:CombinedSchematic}
\end{figure*}

\begin{figure*}[ht]
  \includegraphics[width=\textwidth]{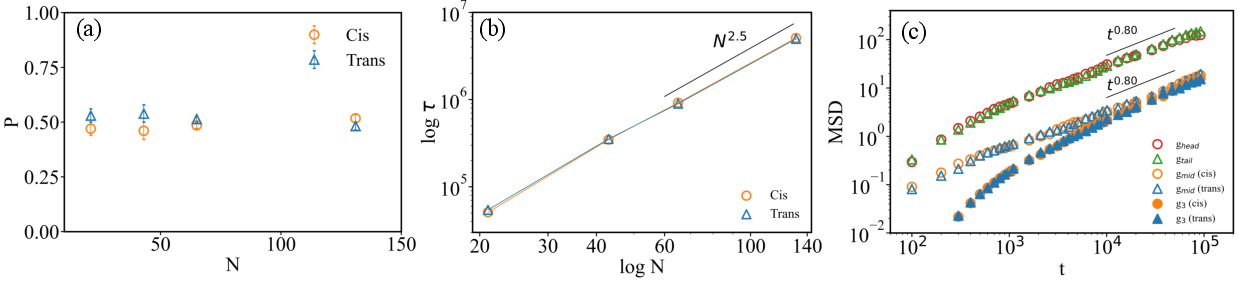}
  \caption{(a) Translocation probability \textit{P} of a free polymer of length $N = 21, ~43, ~65,$ and $131$. {$P_c$} represents the translocation probability to the cis-side (circle) while {$P_t$} shows the translocation probability to the trans side (triangle). (b) Scaling of the translocation time $\tau$ with respect to the polymer of length $N = 21, ~43, ~65,$ and $131$ for no crowding case. Average $\tau$ shows the same behavior in the case of cis and trans translocation for varying lengths of polymer with the scaling of $\tau_c \sim \tau_t \sim N^{2.5}$. (c) The mean-squared displacement of the first monomer $g_{head}(t)$, middle monomer $g_{mid}(t)$, end monomer $g_{tail}(t)$ and center of mass $g_{3}(t)$ of the polymer of $N = 65$. The curve shows alike behavior irrespective of the side (cis/trans), giving $\alpha = 0.8$ in 2D  at long times. }
  \label{fig:free_p_t_MSD}
\end{figure*}

\section{Methods and Modeling}

We studied the unforced translocation process of a flexible self-avoiding polymer in a free and crowded two-dimensional (2D) environment. The polymer of length $L$ is represented by a space curve $\textbf{r}(s)$ in 2D, where the parameter $s$ goes from $s = 0$ to $L$, along the chemical distance or contour length of the chain. The standard two-dimensional coarse-grained bead-spring model of the polymer chain is used to represent the translocating polymer. In the discrete limit, the polymer consists of $N+1$ monomers of size $\sigma$ which are connected by $N$ Finite Extension Nonlinear Elastic (FENE) springs,

\begin{equation}\label{eq:1}
U_{FENE} (l_{ij}) = 
\begin{cases}
 -\frac{1}{2} k l_{0}^2 \ln \left(1 - \frac{l_{ij}^2} {l_{0}^2} \right), & \text{for~} l_{ij} \leq l_{0} \\
                              \infty, & \text{otherwise}.
\end{cases}
\end{equation}
Here $k$ is the FENE spring constant, $l_{ij}$ is the instantaneous separation, and  $l_{0}$ is the maximum allowed length between two consecutive monomers. In the discrete picture, the $i^{th}$ bead is represented by a 2D position vector $\textbf{r}_i(t)$. The excluded volume interaction exclusively between the beads, crowders, and bead-crowders is modeled by a standard truncated short-range repulsive Lennard-Jones (LJ) potential, commonly known as Weeks-Chandler-Andersen (WCA) potential as \cite{weeks1971role}

\begin{equation} \label{eq:2}
 U_{LJ}({r_{ij}}) = 
 \begin{cases}
          4\epsilon \left[\left(\frac{\sigma_{ij}}{r_{ij}}\right)^{12}-\left(\frac{\sigma_{ij}}{r_{ij}}\right)^{6}\right] + \epsilon, \quad & \text{for ~}  r_{ij} \leq 2^{1/6} \sigma_{ij} \\
          0, \quad & \text{otherwise}. 
\end{cases}
\end{equation}

Here $r_{ij}=|\textbf{r}_i-\textbf{r}_j|$ is the distance between $i$ and $j$ particle, $\sigma_{ij}= (\sigma_i + \sigma_j)/2$  is the diameter of the effective interaction region of the interacting pairs with diameter $\sigma_i, \sigma_j$ and $\epsilon$ is interaction strength of the potential. The use of a purely repulsive part of the LJ potential mimics the good solvent by cutting out the attractive effect.

We consider a two-dimensional rectangular box separated by a repulsive wall in the middle consisting of rigid repulsive LJ particles of thickness the same as the diameter of a monomer with bead size  $\sigma = 1$. The middle wall separates the box into two equal regions; the left side is referred to as \textit{cis} while the right side is called \textit{trans}. These two separate regions \textit{cis} and \textit{trans}  are connected by a pore of length $w_H = \sigma$ and pore width \textit{$w_V = 1.6  \sigma$} at the center of the wall \cite{chen2013dynamics}. A study of the effect of box size has been done. The box has been chosen to be big enough so the polymer does not feel any effect from the box size. We have checked the effect of box size for different scaling parameters, such as translocation time and probability, and our model matches with both experimental and analytical theories of scaling regime. The geometry of the cavity plays a crucial role and has been studied extensively \cite{polson2019polymer, sharma2022driven, saltzman2009conformation, palyulin2014polymer, ikonen2012polymer,slonkina2003polymer}. 

In our simulation, the equation of motion of the dynamics of the $i^{th}$ monomer is governed by the Langevin equation by neglecting hydrodynamic interactions among the monomers as 
\begin{equation} \label{eq:3}
\begin{split}
 m \frac{d^2 \textbf{r}_i(t)} {dt^2} &= - \xi \textbf{v}_i(t)+ {\textbf{F}_i^R}(t) - \sum _{j=1, ~i \neq j} ^{N + 1} \boldsymbol{\nabla} U_{LJ} (|\textbf{r}_i - \textbf{r}_{j}|)  \\ 
    & - \boldsymbol{\nabla} U_{FENE} (|\textbf{r}_i - \textbf{r}_{i \pm 1}|) - \sum _{j=1} ^{N_c} \boldsymbol{\nabla}U_{LJ} (|\textbf{r}_i - \textbf{r}_{j}^c|) \\
    & -  \boldsymbol{\nabla}U_{LJ} (|\textbf{r}_i - \textbf{R}_{Wall}|) -  \boldsymbol{\nabla} U_{LJ} (|\textbf{r}_i - \textbf{R}_{Box}|), 
\end{split}
\end{equation}
where $\textbf{r}_i$ is the position, $m$ is the mass, $\textbf{v}_i$ is the velocity of the $i^{th}$ monomer, $N_c$ is the number of crowder, $\textbf{r}_j^c$ is the position of the $j^{th}$ crowding particle and $\xi$ is the friction coefficient, and thermal fluctuation is considered by $F_i^R$ which is the random forces satisfying fluctuation-dissipation theorem, 
\begin{equation}\label{eq:4}
\langle {\textbf{F}}_i^R(t) \rangle =0, \langle{\textbf{F}}_i^R(t) \cdot {\textbf{F}}_j^R(t')\rangle = 2 d k_BT \xi \delta_{ij} \delta(t-t'),
\end{equation}

 that connect the particle diffusivity to the friction coefficient $D=k_BT/\xi$. Here $d$ is the dimension of the system, in our case $d=2$, $k_B$ is the Boltzmann constant, $T$ is the temperature, and $i$ and $j$ represent the coordinate components. The presence of the confining rectangular box is represented by $\textbf{R}_{Box}(x,y)$, where $x\in[-L_x, L_x]$, $y\in [-L_y, L_y]$ and $L_x=2L_y$. The wall in the middle is represented by $\textbf{R}_{Wall} (x=0, y) $ and the pore is represented by $\textbf{R}_{pore} (x=0, y=0) $.

Similarly, the corresponding Langevin equation \cite{luo2006polymer} for the dynamics of the $i^{th}$ crowding particle can be written, where $m_i^c$ is the $i^{th}$ crowding particle's mass, $r_i^c(t)$ is crowder position at time $t$, and other crowding particles at position $r_j^c(t)$, $\xi_i$ is friction coefficient, $v_i^c$ is the $i^{th}$ crowder's velocity. The sizes of a crowder are represented as $\sigma_c$ for the cis-side and $\sigma_t$ for the trans-side. The friction and mass of the crowder on each side are proportional to their sizes, with 
$\frac{\xi_t}{\sigma_t} = \frac{\xi_c}{\sigma_c}$ and  $\frac{m_t}{\sigma_t^2} = \frac{m_c}{\sigma_c^2}$ respectively.

We are using LJ parameters m, $\sigma$, $\epsilon$ for defining mass, distance, and energy scales corresponding to $t=\sqrt{m \sigma^2 / \epsilon} $ and $f= \epsilon / \sigma$ for time scales and force scales in $ps$ and $pN$ respectively.  The dimensionless parameters used in the simulation are temperature $T = 1.2$, the maximum allowed separation between each monomer $R_{0} = 1.5$, spring constant $k = 30$, friction $\xi = 0.7$, the mass of each monomer $m = 1$. 

For each translocation process, as a starting configuration, we placed the middle monomer $(N/2 + 1)$ at the center of the pore. We keep the middle monomer fixed until the rest of the polymer equilibrates for $10^5$ time steps. All the simulations are performed using a Langevin thermostat, to begin with, a truly random initial configuration. The equation of motion is integrated using the velocity Verlet algorithm in each step. After the equilibration time, the middle monomer is released to allow the translocation process. After each successful translocation event, where the polymer ends up moving completely to either side of the wall, the simulation is stopped. In order to get satisfactory statistics, we averaged our data over $10^3$ independent realizations, with a time step $\Delta t = 0.005$.

\begin{figure}[ht!]
        \centering
        \includegraphics[width=\linewidth]{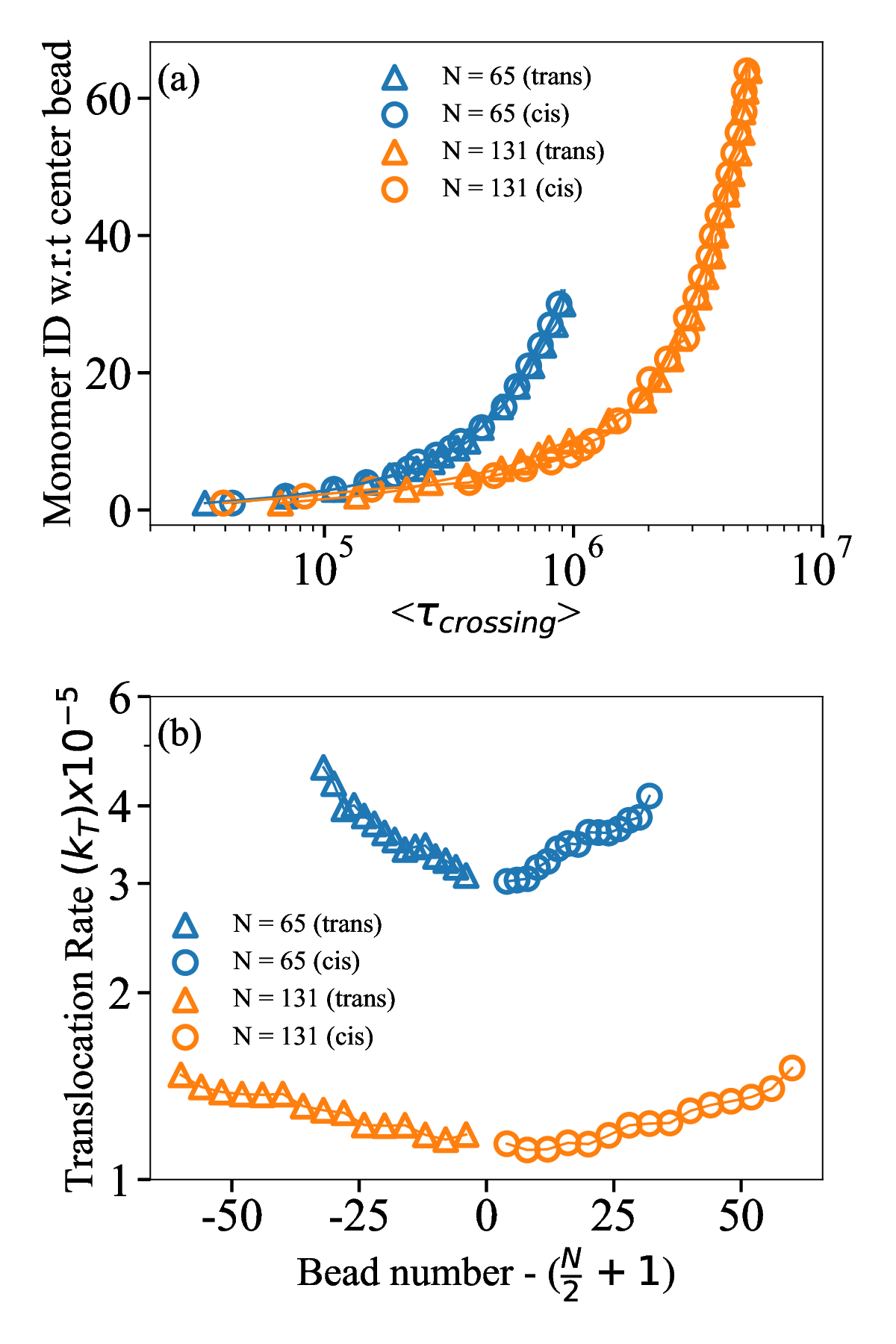}
        \caption{(a) For a polymer chain of $N$ monomers labeled $i = 1, ~2, ~3, \cdots, ~N$ having center at $N/2 + 1$, its left-side can be represented as $j_L = i,  1 \leq i < (N/2 + 1) $ and right-side of the chain as $j_R = (N/2 + 1) - i$,  $(N/2 + 1) \leq i \leq N $. Monomer id w.r.t. center bead $N/2 + 1$ of the polymer against average bead crossing time $\tau_{crossing}$ of polymer length $N = 65$ (blue), $~131 $ (orange) is plotted.  (b) Translocation rate $k_T$ of a polymer of length $N = 65, ~131$. The graph is plotted symmetrically w.r.t to the center bead at the pore. Curves on the left represent the translocation rate of $j_L$ to show trans-translocation (triangles), and the right curve represents the translocation rate of $j_R$ to show cis-translocation (circle).}
        \label{fig:time_of_crossing_rate_N65.eps}
\end{figure}

        \begin{figure}[!ht]
        \centering
        \includegraphics[width=\linewidth]{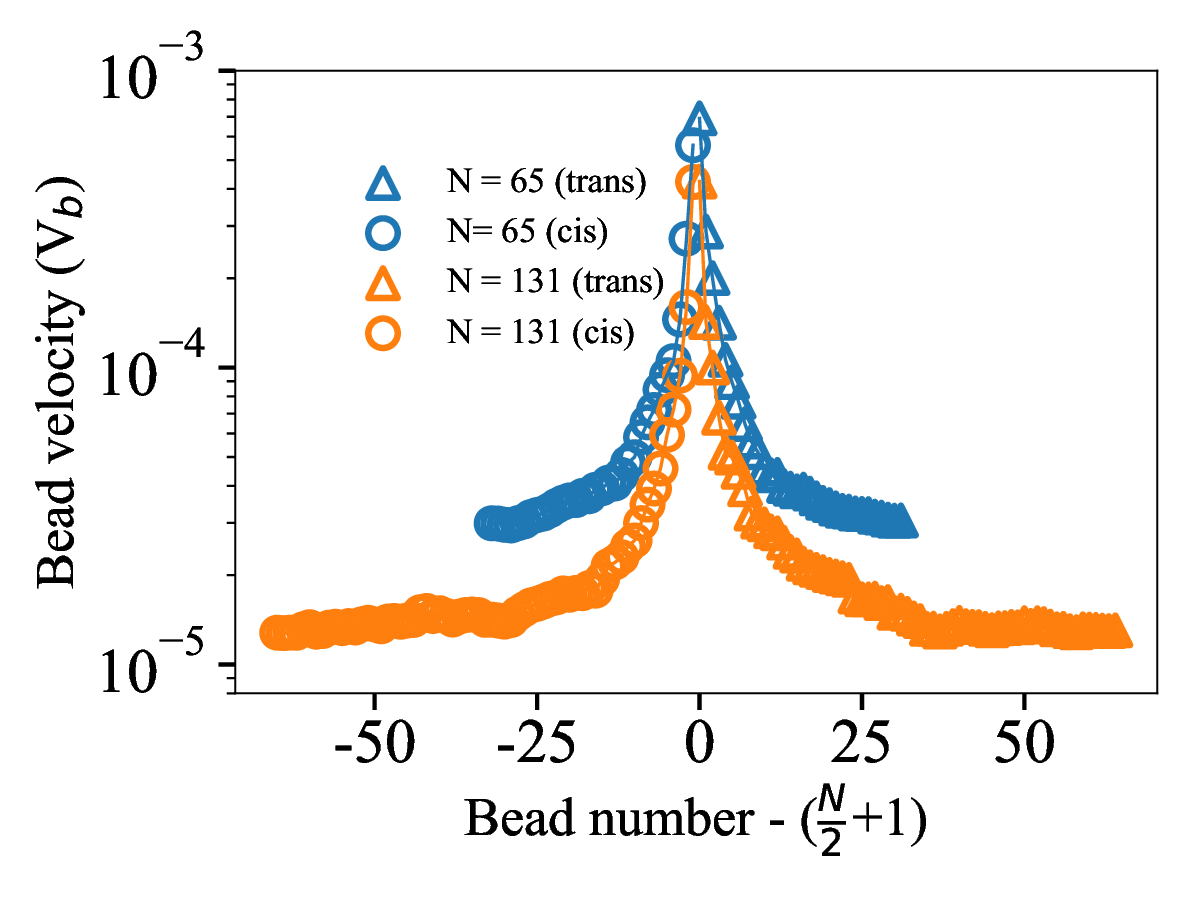}
		%\vspace{20mm}
	\caption{Bead velocity $v_b$ of the individual monomer for a polymer of length $N = 65, ~131$. The graph is plotted symmetrically w.r.t. center bead ($N/2 + 1$) at the pore and indicates bead velocity when moving to the trans-side (triangle) and cis-side (circle).} 
        \label{fig:Trans_location_speed_N65_N131.eps}
    \end{figure}

        \begin{figure}[h!]
        \centering
        \includegraphics[width=\linewidth]{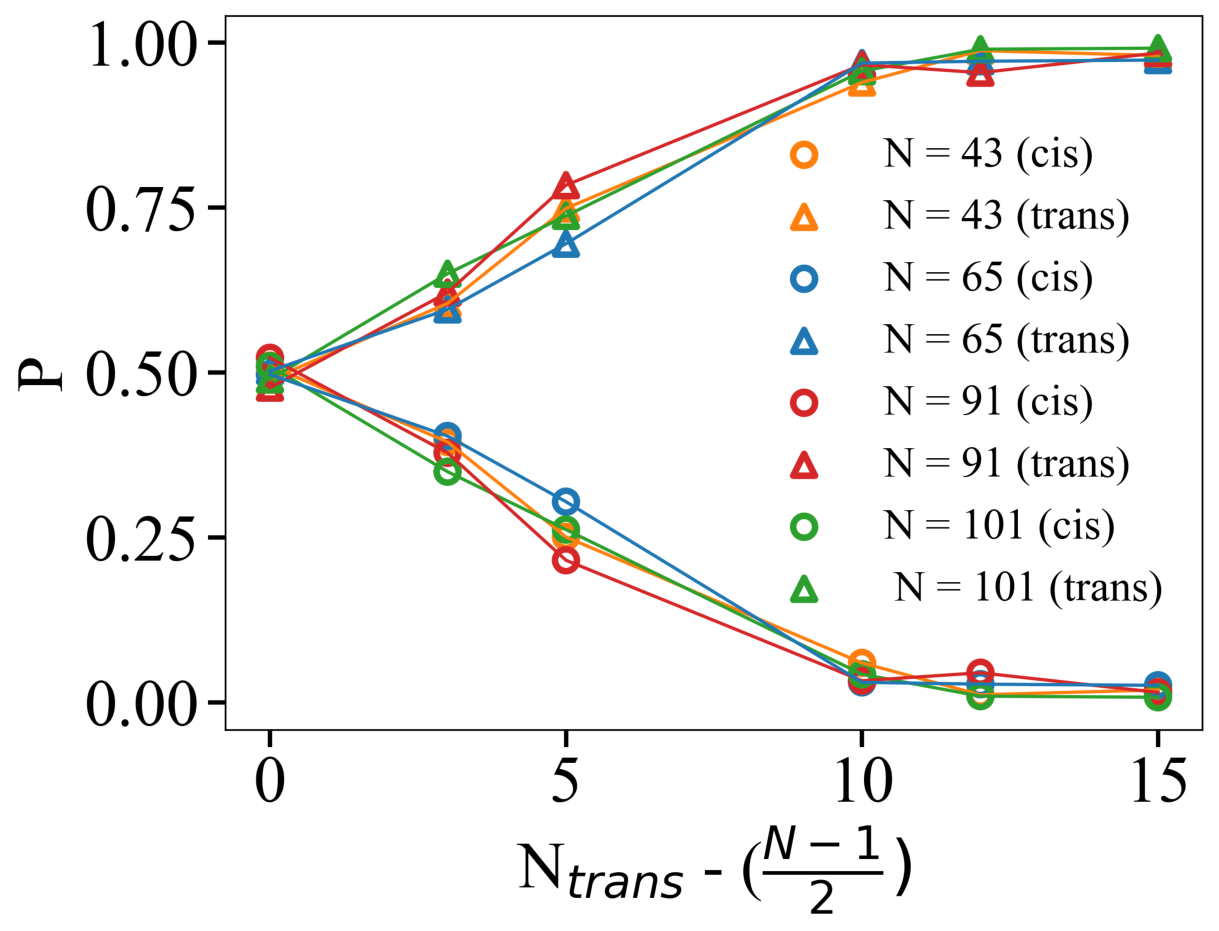}
        \caption{Plot of probability of translocation vs. the number of monomers being shifted to the trans side. Here, $N_{trans}$ is the length of the polymer on the trans side, and $(N-1)/2$ represents the middle of the polymer length $N$. The probability of overcoming the nucleation barrier for further successful translocation for asymmetrically placed polymer towards the trans-side is plotted.  Asymmetry introduced in the polymer, which was initially placed symmetrically at the pore, is represented in terms of the number of monomers shifted from its middle to the trans side and shown as $N_{trans}-(N-1)/2$. For symmetrically placed polymer, $N_{trans}-(N-1)/2 = 0$, which gives $N_{trans} = (N-1)/2$. The probability of cis-side (circle) and trans-side (triangle) for the different $N$ is shown. It starts from $P =0.5$ for the symmetrically placed polymer when $N_{trans} = (N-1)/2$ and reaches saturation when a sufficient amount of monomers has been shifted for the nucleation phenomenon.}
        \label{fig:probabilityofshiftedmonomers}
    \end{figure}

     \begin{figure}[h!]
        \centering
        \includegraphics[width=\linewidth]{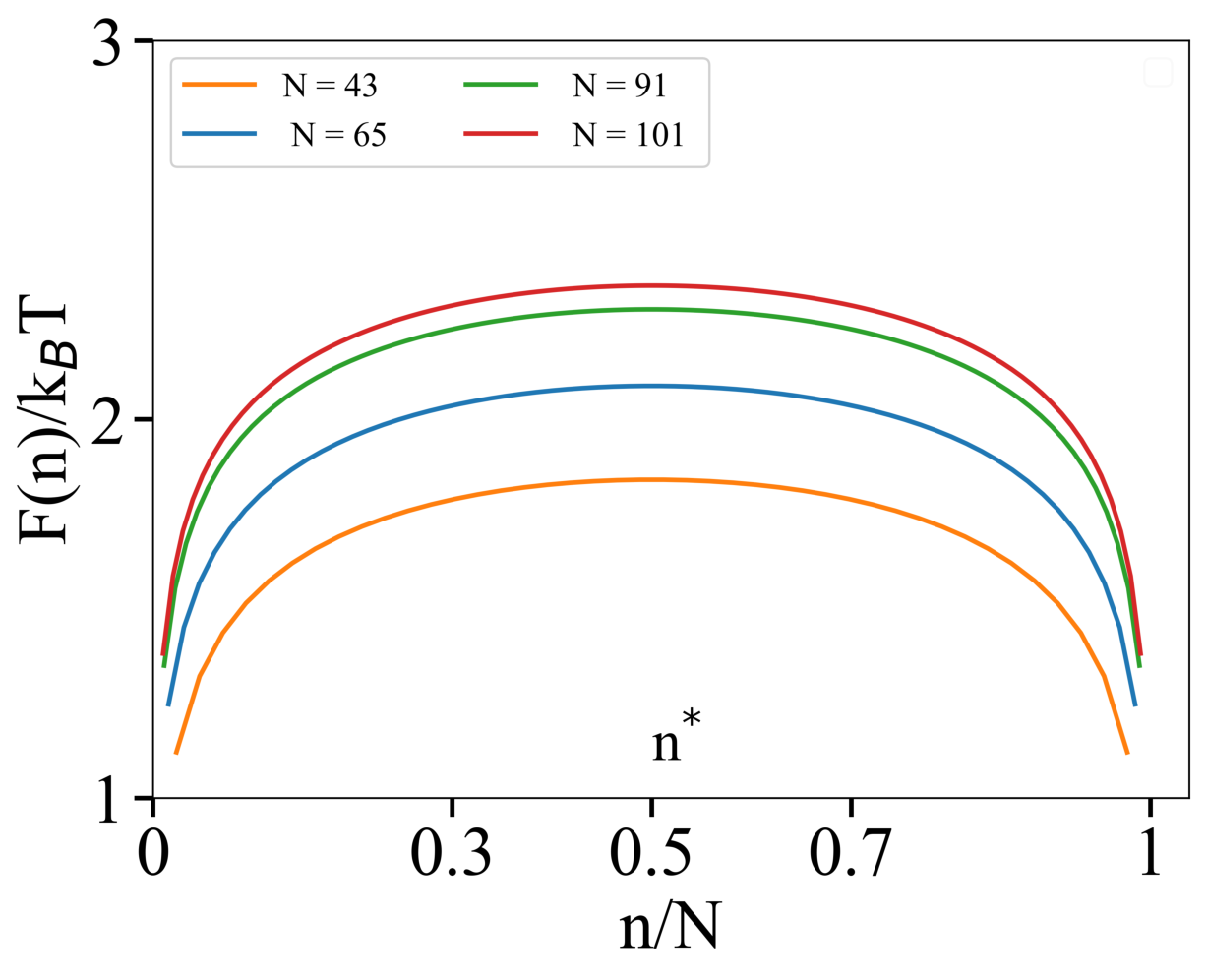}
        \caption{Plot of Free energy $F(n)$ vs. $n/N$, where n represents the number of asymmetrically placed segments, and N is the polymer length. Here $n$ varies from the initial monomer of the polymer chain to its entire length.   $n^{*}$ is the critical number of polymer segments  representing the free energy maxima.}
        \label{fig:symmetricfreeenergy}
    \end{figure}

    \begin{figure}
        \centering
        \includegraphics[width=\linewidth]{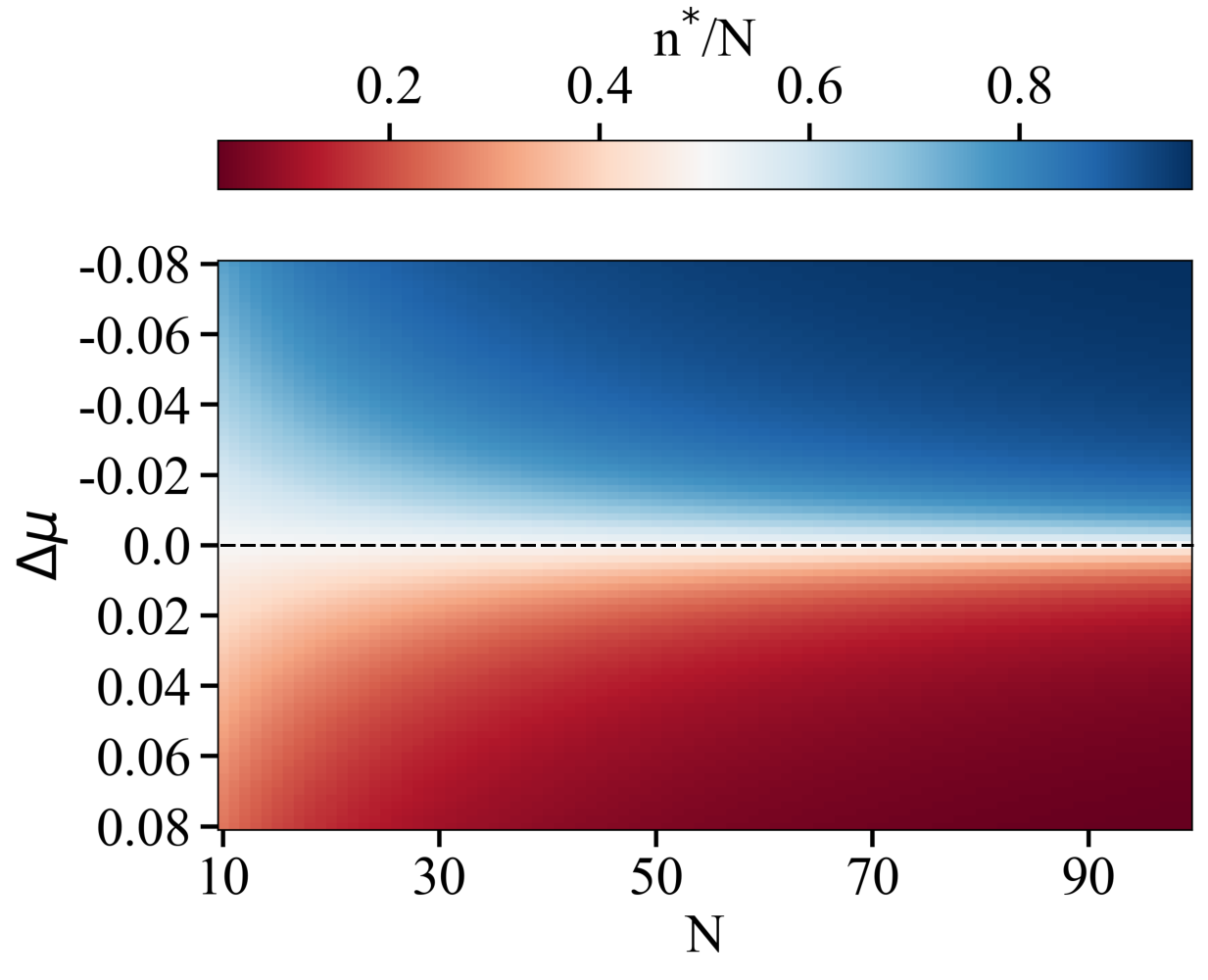}
        \caption{Here, $n^{*}$ is the critical number of polymer segments which need to be translocated for successful translocation event and represent the free energy maxima, $N$ is polymer length and $\Delta \mu$ is the chemical potential gradient between cis and trans-side. The black dashed line represents no crowding case where $\Delta \mu = 0$ and $n^{*}/N$ is exactly half for all chain lengths $N$, and the free energy barrier is symmetric. The chemical potential gradient varies from negative to positive, shown by blue to red. }
        \label{fig:heatmap_chemicalpotential}
    \end{figure}

      \begin{figure}[h!]
        \includegraphics[width=\linewidth]{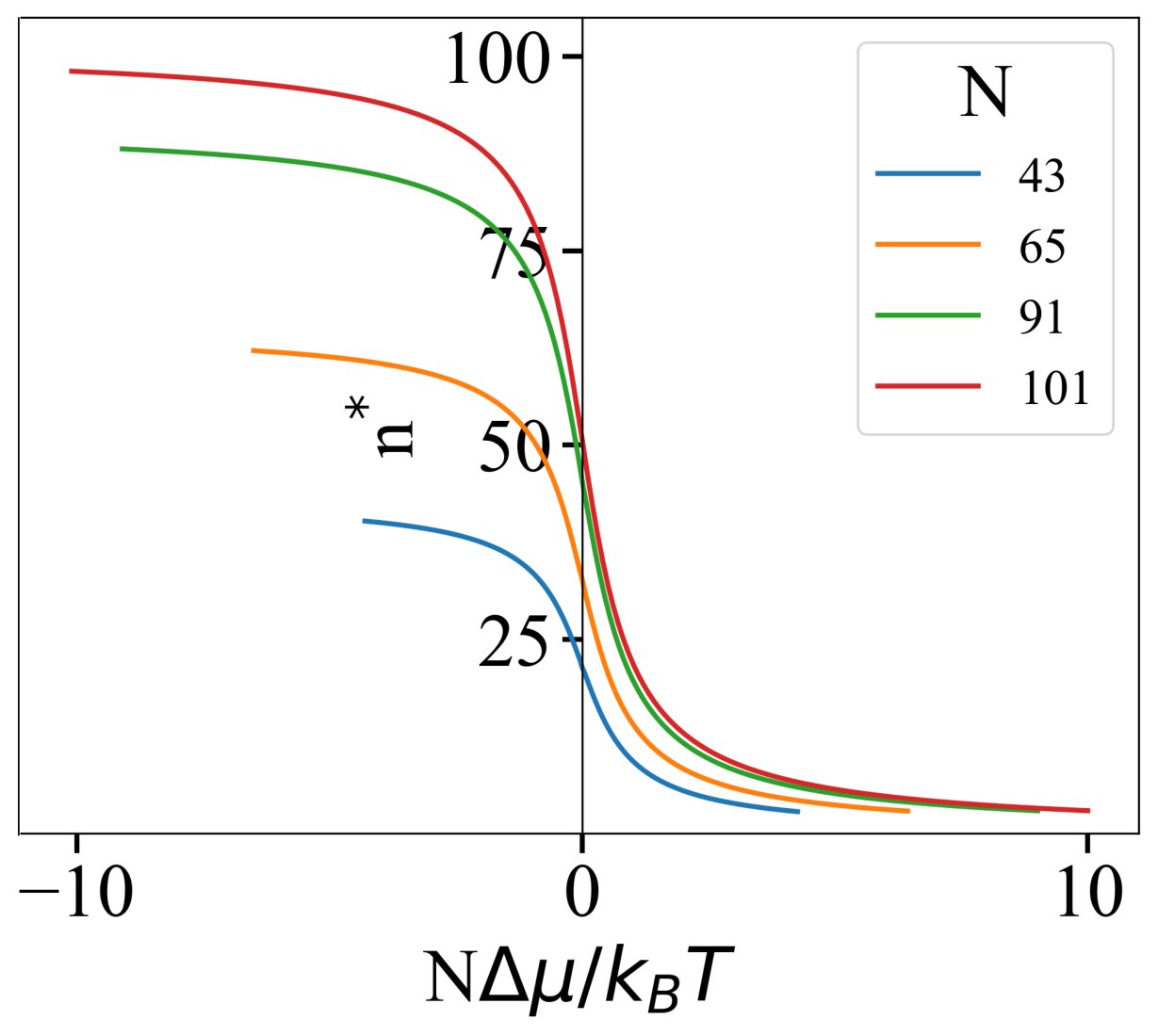}
        \caption{ Change in n$^*$ against $N\Delta\mu$. Plot showing the dependence of the critical number of polymer segments translocated on the trans side $n^{*}$ on  $N\Delta\mu/k_BT$ for the polymer of different lengths $N= 43, ~65, ~91, ~101$. When $\Delta\mu=0$, which leads to n$^*=N/2$ as we move towards the positive side, n$^*$ decreases, and all the N collapses. This graph considers two sides of the wall: the receiver and the donor. The chemical gradient is calculated from the difference between the donor and receiving side, and the sign decides the direction in which translocation could be preferred.}
        \label{fig:freeEnergyVsN}
    \end{figure}

\section{RESULTS AND DISCUSSION}
\subsection{\textbf{Translocation without crowders}}
\subsubsection{Translocation probability \textit{(P)}, time \textit{($\tau$)}, and MSD}

We first considered the polymer translocation in a crowd-free medium (Fig. \ref{fig:CombinedSchematic}a ). We calculated the translocation probability $P$ and time $\tau$. The translocation probability $P$ can be defined as the ratio of successful translocation events towards a particular side to the total number of successful events \cite{lehtola2010unforced, luo2006polymer}. The probability of the polymer to translocate to \textit{cis} (left) side is represented by $P_c$ and to the \textit{trans} (right) side is $P_t$. The time required for a polymer positioned symmetrically with a center bead initially at $t=0$ in the middle of the pore to fully translocate to either side of the wall is referred to as translocation time $\tau$. If the polymer goes to the \textit{cis} (left) side, we call it a \textit{cis}-translocation time $\tau_c$, and for \textit{trans} (right) side, it is a \textit{trans}-translocation time $\tau_t$.
 
 Ideally, in the absence of crowders (Fig. \ref{fig:CombinedSchematic}a), the system remains unbiased from either side and exhibits an equal probability of translocation \cite{lehtola2010unforced}. As expected, from our model, we have observed equal \textit{cis} and \textit{trans} translocation probability, $P_c \approx P_t \approx 0.5$, for different polymer length $N$ (Fig. \ref{fig:free_p_t_MSD}a). The translocation time $\tau$ scales as $\tau_c \sim \tau_t \sim N^{2.5}$ (Fig. \ref{fig:free_p_t_MSD}b), as we vary length $N$, exactly matches with analytical and previously found simulation studies on unforced free polymer translocation \cite{gopinathan2007polymer, chuang2001anomalous}.

To understand the dynamic behavior of the polymer, we analyzed the time evolution of mean squared displacement (MSD) of the center of mass $g_3(t)$ and individual monomers: the first monomer $g_{head}(t)$ for $s=1$, mid monomer $g_{mid}(t)$ for $s=N/2+1$, and end monomer $g_{tail}(t)$ for $s=N+1$. In general, the MSD takes the following power law form: $\langle \Delta r^{2} (t)\rangle \sim t^{\alpha}$, where $\alpha$ is the anomalous diffusion exponent. Each danging end outside the pore has a characteristic relaxation time $\tau_R$ much faster than that of the monomer moving through the pore. The relaxation time (Rouse time) $\tau_R$ of a polymer is typically defined as the characteristic time it takes to diffuse a distance of the order of its size \cite{rubinstein2003polymer}. At small time scales when time $t$ is much less than the relaxation time $\tau_R$ ($t < \tau_R$), the movement of the individual monomers is different from the center of mass movement, resulting in a significant difference in their $MSD(t)$. At a later time, when $t$ is greater than relaxation time $\tau_R$ $(t>\tau_R)$, the MSDs of the middle monomer coincide with the center of mass. For the translocation of a self-avoiding chain in 2D, the exponent can be written as $\alpha = 2/(1+2 \nu)$ where $\nu$ is the flory swelling exponent. In the long time limit, the $g_{mid}(t)$ and $g_3(t)$ converge with $\alpha = 0.8$ for $\nu = 0.75$ (Fig. \ref{fig:free_p_t_MSD}c) \cite{chuang2001anomalous, lehtola2010unforced}. Moreover $g_{head}(t)$ and $g_{tail}(t)$  also show similar behavior with exponent $\alpha = 0.8$. However, as the two ends of the polymer danging in bulk with more translational degree of freedom, the MSD values are always greater than the other or center of mass monomers.

\subsubsection{ Translocation Rate, Crossing Time and,  Bead Velocity }
During translocation processes, it is observed that different segments of the polymer move differently. To envision this interesting phenomenon, we calculated the average bead crossing time $\tau_{crossing}$, translocation rate $k_T$, and bead velocity $v_i(t)$ for different segments of the polymer. Our focus centers on investigating and highlighting the translocation rate and bead velocity of an individual monomer, excluding considerations for the polymer as a whole.

The average time taken by a monomer to translocate through the pore is defined as the average bead crossing time $\tau_{crossing}$. Fig. {\ref{fig:time_of_crossing_rate_N65.eps}a} shows the variation in crossing time taken by each monomer to pass the pore starting from their respective initial position in bulk. As expected, for different monomers, as we move along the backbone of the chain, the crossing times are different. Due to the closeness of the pore, the central monomers take a shorter time to pass the pore over their faraway peers, the tail monomers. In spite of this trivial realization, if we look closely, we can observe that the rate of change of $\langle \tau_{crossing} \rangle$ with respect to monomer id is not constant but rather increases as it moves towards the tail. This interesting feature will be addressed during the discussion of $k_T$ in the following section.

The translocation rate $k_T$ is defined as the number of beads passing through the pore per unit time. $k_T$ is calculated by taking the derivative of the curve in Fig. \ref{fig:time_of_crossing_rate_N65.eps}a. As the translocation process starts with the center bead $(N/2 +1)$ at the pore, the translocation rate increases slowly for beads near the center of the polymer chain due to alike entropic force from both sides. The small values of $k_T$ in the middle clearly indicate that during the translocation event, the monomers in the middle region have several back-and-forth movements around the pore before they fully translocate to one side. Subsequently, once the translocation of some significant middle portion of the polymer has taken place, follow-up monomers motion is governed by the already translocated part, leading to a higher translocation rate for the tail monomers (Fig. \ref{fig:time_of_crossing_rate_N65.eps}b). 
The ratio of the translocation rates of the center monomer with respect to the polymer end monomer are defined as $\overline{k}_L = {k_T (n_{first})} / {k_T (n_{center})}$ for the cis-side, $\overline{k}_R = {k_T (n_{end})} / {k_T (n_{center})}$ for the trans-side, and the average translocation rate $\langle k_0 \rangle = (\overline{k}_L + \overline{k}_R) / 2$. The average translocation rates, $\langle k_0 \rangle$, for the initial and end monomers on the cis ($\overline{k}_L$) and trans ($\overline{k}_R$) sides relative to the central monomer exhibit almost similar ratios of approximately 1.4 and 1.3 for lengths $N = 65, 131$, respectively (see Table \ref{tab:translocation rate ratios}).

\begin{table}
    \centering
%\begin{tabular}{|Q[c,m,1.8cm]|Q[c,m,1.8cm]|Q[c,m,1.8cm]| Q[c,m,1.8cm]|}
\begin{tabular}{|c|c|c|c|}
\hline
 $~~~~~ N ~~~~~$ & $~~~~~ \overline{k}_L ~~~~~$ & $~~~~~ \overline{k}_R ~~~~~$ &  $~~~~~ \langle k_0 \rangle ~~~~~$ \\ \hline \hline
         65 &  1.37 & 1.49 & $\sim{1.4}$   \\   \hline
        131 & 1.32  & 1.25 & $\sim{1.3}$  \\   \hline
\hline
\end{tabular}
   \caption{Motion of the center monomer with respect to the polymer end monomers on the cis-side and trans-side is depicted as the ratio of translocation rate and represented as  $\overline{k}_L$ and $\overline{k}_R$  respectively. The average translocation rate, $\langle k_0 \rangle$, is the mean of $\overline{k}_L$ and $\overline{k}_R$. We observe a similar average translocation rate ratio for different polymer lengths $N$.}
    \label{tab:translocation rate ratios}
\end{table}

Now let us look at the individual bead velocity $v_{i}(t)=dr_{i}(t)/dt$, which tracks the dynamics of the movement of the $i^{th}$ bead during a translocation process. We have observed that middle monomers move faster than tail monomers (Fig. \ref{fig:free_p_t_MSD}c). Being close to the pore, the monomers in the central region are restricted to exhibit quasi-one-dimensional diffusion from the beginning of the translocation process until a successful crossover of that bead happens. While for most of the translocation time, the tail monomers undergo 2D diffusion until they approach the vicinity of the pore, resulting in lower $v(t)$. From the MSD Study in Fig. \ref{fig:free_p_t_MSD}c and bead velocity in  Fig. \ref{fig:Trans_location_speed_N65_N131.eps}c, it can be observed that initially, the dangling tail monomers wander in the medium for a reasonable amount of time and diffuses slowly. In contrast,  the middle monomers in the vicinity of the pore, part of quasi-one-dimensional diffusion, diffuses faster, supporting the notion that resulted in higher $v(t)$ for middle monomers than the tail monomers during the translocation process \cite{afrasiabian2020journey, rezaie2022translocation}. The monomer velocity dramatically increases when it approaches the vicinity of the pore. 

\subsubsection{Asymmetrically placed polymer}

Until now, in the free environment, we have studied translocation by always placing the polymer symmetrically between the cis and trans side, with the middle monomer at the pore as the initial configuration. Now, we are breaking the symmetry by asymmetrically placing the polymer from its middle with extra length into the trans side. As a result, we are designing the initial configuration so that when the translocation process starts, the polymer portion on the trans side is larger than the cis side. We want to know what the free energy of the chain is as the polymer translocates from one region to another.

We are interested in examining the impact of the asymmetrically placed polymer on translocation probability see Fig. \ref{fig:probabilityofshiftedmonomers}. When a polymer translocates through the pore, the chain goes through a decrease in its conformational entropy, and the excluded volume interactions become more pronounced, increasing the free energy of the chain. Here, pore size is assumed to be short in comparison with the length of polymer molecules that can be considered as a small hole in an infinite wall and can allow only one monomer to pass through at a time \cite{sung1996polymer, muthukumar1999polymer, di1997phase, lehtola2010unforced}. 
The two translocating polymer segments on the cis and trans side can be treated as two thermodynamic ensembles separated by the wall. At any state, let there be $n$ segments of the translocating polymer in the trans-side referred to as {\it reciever} and $N-n$ segments in the cis-side as {\it donor}, both are in separate thermal equilibrium. The terms "donor" and "receiver" designate the sides for polymer translocation, with the donor being where the polymer originates and the receiver being the target (translocated) side. In our study, we named the cis side as the donor and the trans side as the receiver for clarity. In the context of observing free crowders, side nomenclature can be chosen for convenience.
The partition sum $Z$ consisting of the total number of conformations of the chain can be written as 
%\begin{equation}
$ Z = Z_{d}(N-n) ~Z_{r}(n).  $ 
%\end{equation}
The partition sum $Z(N)$ for a long tail $N >> 1$ of $N$ segments in the semi-infinite space (half) bounded by an impenetrable wall to which one end is anchored is given as 
%\begin{equation}
    $Z_{half}(N) = \bar{z}^N  N^{\gamma -1}$
%\end{equation}
where $\bar{z}$ is the effective coordination number for the orientation of adjacent bonds and is commonly known as the connective constant. It can be alternatively written as $\exp(-\mu/k_B T)$, where $\mu$ is the chemical potential per segment. $\gamma$ is the critical exponent and depends on the nature of the polymer and the solution. To mimic a good solution condition for a self-avoiding polymeric chain, we have $\gamma$= 0.69. The Helmholtz free energy is  $F_{half}(N)= -k_B T ln Z_{half}(N),$ elaboratively can be written as 
%\begin{equation}
    $\frac{F_{half}(N)}{k_B T}= \frac{\mu N}{k_B T}+(1-\gamma)\ln N.$
%\end{equation}
 The logarithmic part plays a significant role in establishing a free energy profile for polymer translocation. The total free energy of the chain $F(n)$, with $n$ segments translocated into the {\it receiver}-side can be written as the sum of free energies of the two tails, 
\begin{equation}\label{eq:5}
\frac{F(n)}{k_B T}= (1-\gamma_{d})ln(N-n)+ (1-\gamma_{r})ln(n)-\frac{n\Delta\mu}{k_BT},
\end{equation}
where $\gamma_{d}$ and $\gamma_{r}$ are critical exponents in donor and receiver region respectively. $\Delta \mu = \mu_{d} - \mu_{r}$, where $\mu_{d}$ and $\mu_{r}$ are the chemical potentials of the polymer segments in the donor and receiver region respectively. The first two terms on the RHS emerged from the entropy of two tails and clearly resulted in the free energy barrier. In general, for this free energy barrier with its maximum value $F^{*}$, there exists a critical number of translocated segments to {\it receiver} side, that is, $n^{*}$ which can be obtained as the solution of $\partial F(n) / \partial n = 0$, gives
\begin{equation}\label{eq:6}
\frac{n^{*}}{N}= \frac{(\Tilde{\mu}+2 -\gamma_{d}-\gamma_{r})}{2\Tilde{\mu}} 
     -\frac{\sqrt{(\Tilde{\mu}+2-\gamma_{d}-\gamma_{r})^2-4\Tilde{\mu}(1-\gamma_{r})}}{2\Tilde{\mu}}
\end{equation}
where $\Tilde{\mu}= N\Delta \mu/k_B T$. For the crowd-free case, the critical value of $n^{*}$ can be obtained by putting $\Delta \mu =0$ in Equation 11. In our model, we are taking $\gamma_{d} = \gamma_{r} = \gamma$. Then, Equation 9 can be written as 
\begin{equation}\label{eq:7}
\frac{F(n)}{k_B T}= (1-\gamma)ln[n(N-n)].
\end{equation}
The critical number of translocated segments to {\it receiver} side $n^{*}$ for $F^{*}$ is
\begin{equation}\label{eq:8}
\frac{\partial F(n)}{\partial n} = \frac{(1-\gamma)(N-2n)}{n(N-n)},
\end{equation}
solution of $\partial F(n) / \partial n = 0$ gives $n^{*}=N/2$. Fig. \ref{fig:symmetricfreeenergy} shows $F(n)$ vs. $n$, which represents the free energy landscape of polymer translocation in the crowd-free case as a function of the translocation extent. This can be interpreted as when there is a slight deviation from the maxima of $F^{*}$ at $n^{*}= N/2$, the free energy landscape is downhill on both sides (Fig. \ref{fig:symmetricfreeenergy}). The free energy barrier is symmetric when a polymer is symmetrically placed in its middle at the pore. Polymer tends to minimize its energy and is favorable to translocate to either side. In general, the presence of $n^{*}$ indicates that translocation resembles a nucleation phenomenon \cite{muthukumar1999polymer}. If the number of segments is less than $n^{*}$,  then segments still tend to return to {\it donor} side. The translocation process is stochastic, like a nucleation phenomenon. Once a sufficient number of monomers have crossed the nucleation barrier to the {\it receiver} side and are larger than $n^*$, translocation to this side is more favorable as the free-energy profile is now downhill. Hence, a critical number of $n^{*}$ ought to be nucleated in the {\it receiver} side to make translocation successful. The variation of free energy $F(n)$ with the entire length of the polymer with its critical points represented with circles. This free energy curve is helpful in understanding Fig. \ref{fig:probabilityofshiftedmonomers}. In this plot, the probability of overcoming the nucleation barrier for further successful translocation for asymmetrically placed polymer vs. the number of monomers being shifted to the trans side is plotted. Here, $N_{trans}$ is the length of the polymer on the trans side, and $(N-1)/2$ represents the middle of the polymer $N$. Asymmetry introduced in the polymer, which was initially placed symmetrically at the pore, is represented in terms of the number of monomers shifted from its middle to the trans side and shown as $N_{trans}-(N-1)/2$. From the probability plot, it can be seen that as soon as we start placing the polymer asymmetrically on the trans side, that is when it deviates from $N_{trans}-(N-1)/2 = 0$, which gives $N_{trans} = (N-1)/2$ symmetrically placed polymer, it starts preferring the trans (receiver) side.  Once a significant section of the polymer is shifted to the trans (receiver) side (say approximately $60 \%$ of the length) and once the nucleation barrier is crossed, the polymer slowly reaches a free energy minima and it can undergo further translocation to receiver side without coming back to the donor side. 

In Fig. \ref{fig:heatmap_chemicalpotential} we construct phase plots for  $n^{*}/N$ in the $\Delta \mu - N$ plane. The effect of the chemical potential gradient can be seen in Fig. \ref{fig:heatmap_chemicalpotential}. For a crowd-free environment, $\Delta \mu = 0$,  $n^{*}/N$ is exactly half for all chain lengths $N$, and the free energy barrier is symmetric,  shown by the black dotted line in Fig. \ref{fig:heatmap_chemicalpotential}. 
Another situation is translocation into a crowded environment driven by a chemical potential gradient. For the case of one side crowding where the trans side has crowders having $\phi_c = 0$ and $\Delta \mu = \mu_{cis} - \mu_{trans} = 0-\mu =  -\mu$, where $\mu$ is a positive quantity and the polymer tends to move to the free cis side. However, a critical value of polymer segment $n^*$ can reverse this trend, which is sufficient to overcome the osmotic pressure. See the upper panel of Fig. \ref{fig:heatmap_chemicalpotential} and Fig. \ref{fig:freeEnergyVsN}. The free energy barrier is bigger when the chemical potential gradient $\Delta \mu = -\mu$ is in the opposite direction to the translocation process. Since the barrier is bigger despite being crossed, the chain in the receiver region remains in a metastable state and will eventually tend to revert to the donor region.
 For this case, a higher value of  $n^{*}/N$ is needed for a favorable translocation in the receiver region. For lower pannel of  Fig. \ref{fig:heatmap_chemicalpotential} and large values of $N$  and $\Delta \mu$, $n^{*}$ becomes progressively small. We require only a small segment of the polymer to be shifted at the cost of chemical potentials, such as for $N =100$, $\Delta \mu = 0.01$, it can be compensated by $n^{*}=22$ for the nucleation process to happen \cite{muthukumar2016polymer}. In our model of polymer translocating through the crowd-free environment, we have observed that nucleation phenomenons start at $n^{*}=10$. The dependence of Free energy maxima $F^*$ on the chain length $N$ and chemical potential gradient $\Delta \mu$ .

In Fig. \ref{fig:freeEnergyVsN}, we can see a sudden drop in the curve when $\Delta \mu$ deviates from zero. As soon as the crowding is introduced, starting from the small size of the crowders, there is a critical value of $\sigma$ for a change in $\Delta \mu$ at which the polymer chain will face energy downhill with respect to $n$. This trend can be seen in the one-sided crowding case (Fig. \ref{fig:prob_oneside.eps}). For small $\sigma_c$ we have a high value of $\Delta \mu$, crowding has a strong effect on pushing the polymer to the crowd-free side and resulting in a sudden rise which gives $P_c = 1$. While the case of higher $\sigma_c$ and comparatively lower $\Delta \mu$ will create larger voids in the crowding side. The polymer still has a small but non-zero $P_c$. This would suggest the critical point where the translocation can act as the nucleation phenomenon, with few monomers passing through the pore. Very less shifting of polymer, even with the small value of $n$, will escalate the translocation to the crowded side. A detailed discussion of the effect of one-sided (\textit{trans}) crowders on the translocation probability is done in the next section. The dependence of this critical number $n*$ per unit length $N$ with changing $\Delta \mu$ .

\subsection{\textbf{Effect of one-side crowders}} 

\begin{figure*}
        \centering
        \includegraphics[width=\linewidth]{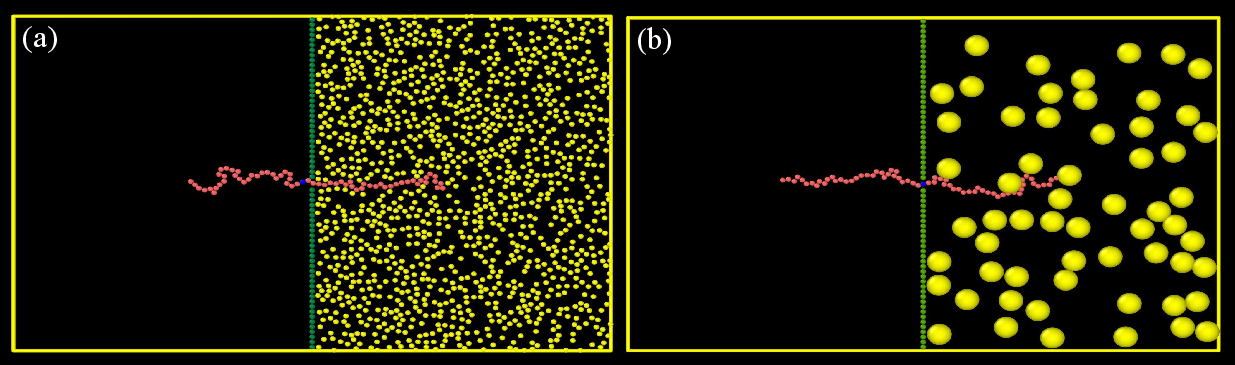}
        \caption{ Schematic illustration of translocation process of a polymer in the presence of crowder on the trans side. (a) The packing fraction of the crowder on the trans side $\phi_t = 0.3$ for $\sigma_t$ = 0.6 (case: $\sigma_t < 1$) (b)Packing fraction of the crowder on the trans side is $\phi_t = 0.3$ for $\sigma_t = 4$ (case: $\sigma_t > 1$).}
        \label{fig:oneide_crowder.eps}
    \end{figure*}

\begin{figure*}
  \includegraphics[width=\textwidth]{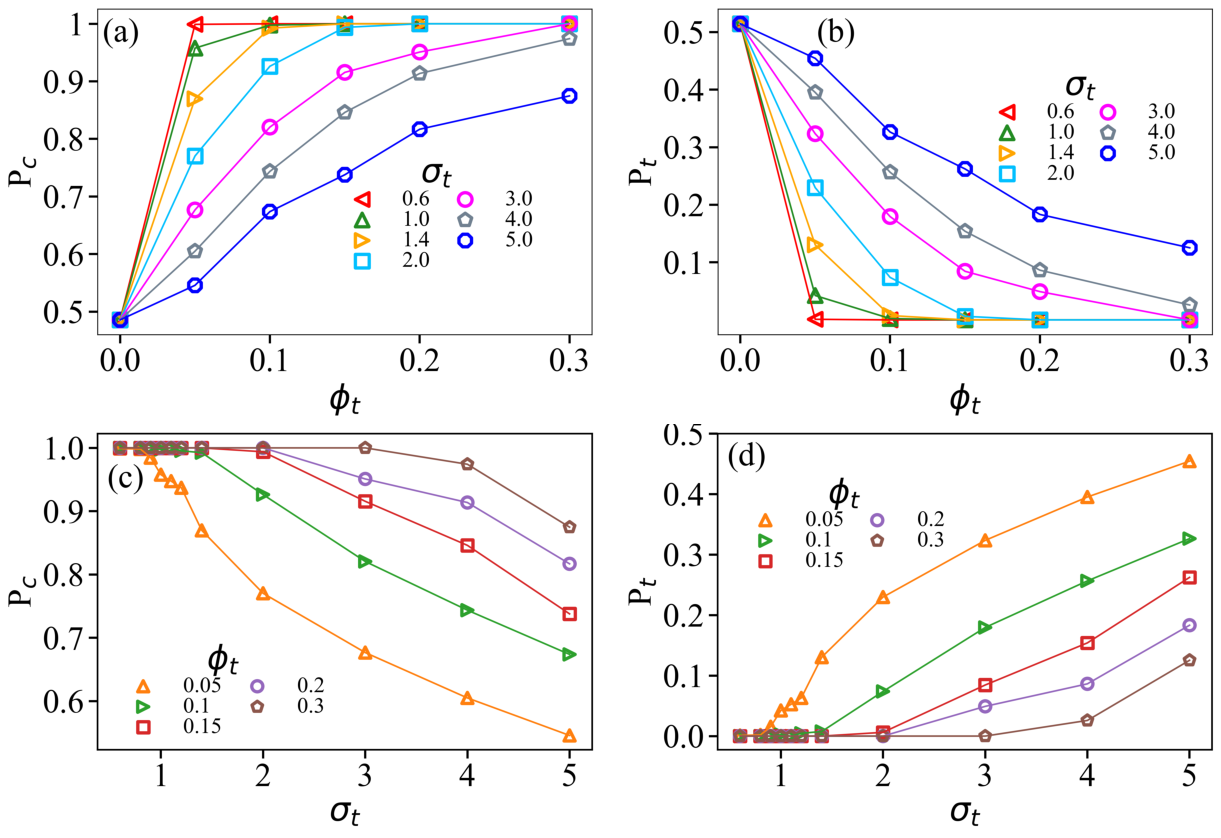}
  \caption{Upper panel: Translocation probability as a function of packing fraction of inert crowders on the trans side $\phi_t$. (a) Translocation probability $P_c$ to cis-side for different $\sigma_t$. (b) Translocation probability $P_t$ to trans-side for different $\sigma_t$. 
  Lower panel: Translocation probability as a function of $\sigma_t$ of inert crowders on the trans side. (c) Translocation probability $P_c$ to cis-side for different $\phi_t$. (d)  Translocation probability $P_t$ to trans-side for different $\phi_t$.}
  \label{fig:prob_oneside.eps}
\end{figure*}

We now study the impact of one-sided inert crowders on the translocation process (Fig. \ref{fig:CombinedSchematic}b). The study of inert crowders on both sides \cite{chen2013dynamics} and non-inert crowders on one side (chaperones) \cite{yu2011chaperone} has been studied. However, a detailed investigation of translocation processes with inert crowders on one side did not get its due attention. For this purpose, we choose to put inert crowders only on one side, say the {\it trans} side, in our case (Fig. \ref{fig:oneide_crowder.eps}) and studied the translocation phenomena of a symmetrically placed flexible polymer. As expected, we have observed that the translocation probability and time have a strong dependency on the packing fraction $\phi_t$ and crowding size $\sigma_t$. To understand the underlying mechanism of this effect, we first quantify the ballpark average distance between the crowders $\langle r_s \rangle $ and mean entropic force $ \langle F_r \rangle $.

Consider $N_t$ is the total number of crowders on the trans-side, each with size, $\sigma_t$ \cite{chen2013dynamics}. The packing fraction on the trans-side, with dimension $L_x \times L_y$ is 
\begin{equation} \label{eq:9}
     \phi = \frac{N_t ~ \pi (\frac{\sigma_t}{2})^2}{L_x \times L_y}.
\end{equation}

Note that the dimension of the box is $2L_x \times L_y$. Now for a fixed $\phi$, the average distance between crowders, taken from their  center $\langle r_c \rangle$ can be written as,
\begin{equation} \label{eq:10}
     \langle r_c \rangle = \frac{\sqrt{L_x \times L_y}}{\sqrt{N_t}}.
\end{equation}
%\begin{math} \label{eq2}
     %\langle r \rangle = \frac{\sqrt{L_x \times L_y}}{\sqrt{N_t}}
%\end{math}.
Hence, the average distance between the surface of crowders $\langle r_s \rangle$ is\\
\begin{equation} \label{eq:11}
     \langle r_s \rangle = \langle r_c \rangle  - \sigma_t = \frac{\sqrt{L_x \times L_y}}{\sqrt{N_t}} - \sigma_t.
\end{equation}

For the box of the same dimension, inserting the value Eq. \ref{eq:9} in the Eq. \ref{eq:11}, we get\\
\begin{equation} \label{eq:12}
    \langle r_s \rangle = \sigma_t \left( \frac{1}{2} \sqrt{\pi/\phi} - 1 \right).  
\end{equation}
 
 From the above expression we observed that the average distance between the surface of the crowders $ \langle r_s \rangle $, that is the space available for the polymer to explore, is proportional to the crowder size, $ \langle r_s \rangle \propto \sigma_t $,  and inversely proportional to the square-root of the packing fraction, $ \langle r_s \rangle \propto 1/\sqrt{\phi}$. This behavior plays a crucial role in understanding the effect of crowders on translocation. A detailed explanation of its significance can be found in the coming section of translocation probability.

The focus of our study is on the regime of packing fraction of crowder $\phi < 0.5 $. According to the percolation theory, the fluid phase is considered continuous  \cite{stauffer2018introduction,essam1980percolation}, indicating the formation of channels by crowders.

Now, we consider average radial entropic force $\langle F_r \rangle$ and its role in the translocation process \cite{muthukumar1989effects,roos2014entropic}. Radial entropic force describes the force exerted on a monomer by the particles as it moves through the channel formed by the crowders and drives the polymer to translocate through the pore. It is proportional to the temperature, inversely proportional to the confinement created by the crowders, and equivalent to inter-crowders separation \cite{sheng2012ejection,chen2013dynamics}. Mean radial entropic force can be written as

 \begin{equation} \label{eq:13}
     \langle F_r \rangle = F_0 \frac{ k_B T}{\langle r_s \rangle} ,
 \end{equation}
 
 where $k_B$ is the Boltzmann constant, $T$ is the absolute temperature, and $F_0$ is the proportionality constant depends on the solvent-polymer interactions. Simplifying the expression in the terms of $\sigma$, $\phi$, we have
 
 \begin{equation} \label{eq:14}
     \langle F_r \rangle = F_0 \frac{k_BT}{\sigma (\frac{1}{2}\sqrt{\pi / \phi}-1)} 
 \end{equation}
 
The presence of crowders on the {\it trans}-side restricts the motion of the polymer and reduces its possible configurational entropy, which in turn reduces the entropic force acting on the polymer. This resulted in a lower entropic state in the {\it trans}-side compared to the {\it cis}-side, which has no crowders and, therefore, less restriction on the polymer's motion. The difference in entropy between the two sides serves as a driving force for the polymer to translocate from the lower entropy ({\it trans} side) to the higher entropy ({\it cis} side) \cite{chen2013dynamics}.

To elucidate the dynamics of polymer translocation through a crowded environment driven by different sizes, we examine the role of osmotic pressure $(\Pi)$. Being a colligative property, \textit{osmotic} pressure is proportional to the solute concentration $\rho_c$, and temperature $T$, can be written as $\Pi=  R \rho_c T$, where $R$ is the constant of proportionality. It can be used to determine the direction of polymer movement through the pore in response to the difference in crowding concentration gradient on either side\cite{doi1988theory, de1979scaling, muthukumar2016polymer}. The polymer is influenced by this osmotic pressure difference, which drives it to the {\it cis}-side where the concentration and, hence, pressure are lower and have more free volume. The entropic force also pulls the polymer in the direction of higher entropy, which is towards the {\it cis}-side. The interplay between these two forces determines the overall direction of polymer translocation through the pore. 

\subsubsection{Translocation probability as a function of $\phi_t$, $\sigma_t$}

Translocation probability to the {\it trans}-side \textit{$P_t$} is defined as the ratio of successful translocation events towards the {\it trans}-side (right) to the total number of successful translocation events that have occurred. The probability of the polymer moving to the free  {\it cis}-side (left) is denoted by \textit{$P_c$} and note for this study of one-sided crowding, we always set  $\phi_c=0$, representing the free environment.

Fig. \ref{fig:prob_oneside.eps}a shows the probability of the polymer to translocate to the {\it cis}-side ($P_c$) as a function of the packing fraction $\phi_t$, for different values of $\sigma_t$. With increasing $\phi_t$, $P_c$ rapidly increases ($P_c=P_t = 0.5$ at $\phi_c=\phi_t$=0) and eventually approaches saturation ($P_c= 1$). This indicates that, for the difference in crowding concentration on either side, the polymer prefers moving to the free {\it cis}-side. It can be understood on the basis of entropic ground and osmotic pressure. {\it Cis}-side has a more free volume, which drives the polymer to translocate to the free side to achieve a higher entropy state by minimizing the crowding-induced constrained configurations. Additionally, the concentration gradient of the crowders also plays an important role in pushing the polymer from the crowded {\it trans}-side to the free cis-side. Whereas for $P_t$ varying with $\phi_t$ at a fixed $\sigma_t$ shows complementary decrease ($P_t = 1- P_c $) in its value (Fig. \ref{fig:prob_oneside.eps}b). This also manifests the same effect of entropy and osmotic pressure.

In Fig. \ref{fig:prob_oneside.eps}c, d we represent how $P_c$ and $P_t$ varies with $\sigma_t$ for different values of $\phi_t$. We observe that $P_c$ decreases and $P_t$ increases monotonically with increasing $\sigma_t$ for different $\phi_t$. This trend continues for the lowest $\phi_t$ until it reaches the equal probable case $P_t \approx P_c \approx 0.5$ at maximum $\sigma_t$. While with increasing $\phi_t$, and for lower values of $\sigma_t$, the curve rises slowly. Correspondingly for the highest $\phi_t$, $P_t$ starts increasing only after $\sigma_t>3$, indicating a need for a higher $\sigma_t$  to reach $P_t \neq 0$ at higher $\phi_t$. The reason is that for a fixed $\phi_t$ on increasing $\sigma_t$, the number of crowders decreases, thereby leading to higher $\langle r_s \rangle$ between the randomly distributed crowders, resulting in the existence of large voids on the {\it trans} side that a polymer can explore and thereby increasing the tendency of the polymer to translocate to the crowders side. Additionally, this behavior simply corresponds to having effective osmotic pressure from the crowders on the {\it trans}-side. The concentration gradient created by crowders from the {\it trans}-side to the {\it cis}-side, that is, from higher osmotic pressure to the lower one, also drives the polymer to move to the free {\it cis}-side. For the highest packing fraction of the crowders $\phi = 0.3$, with increasing $\sigma_t$, crowders have more large voids resulting in more free volume for the polymer to explore and also osmotic pressure at the {\it trans}-side is comparatively less by higher $\sigma_t$ (lower $\rho_c$) than that by lower $\sigma_t$ (higher $\rho_c$) and polymer prefers to stay towards the crowder side which is reflected in $P_t$ curve. 

\begin{figure*}
  \includegraphics[width=\textwidth]{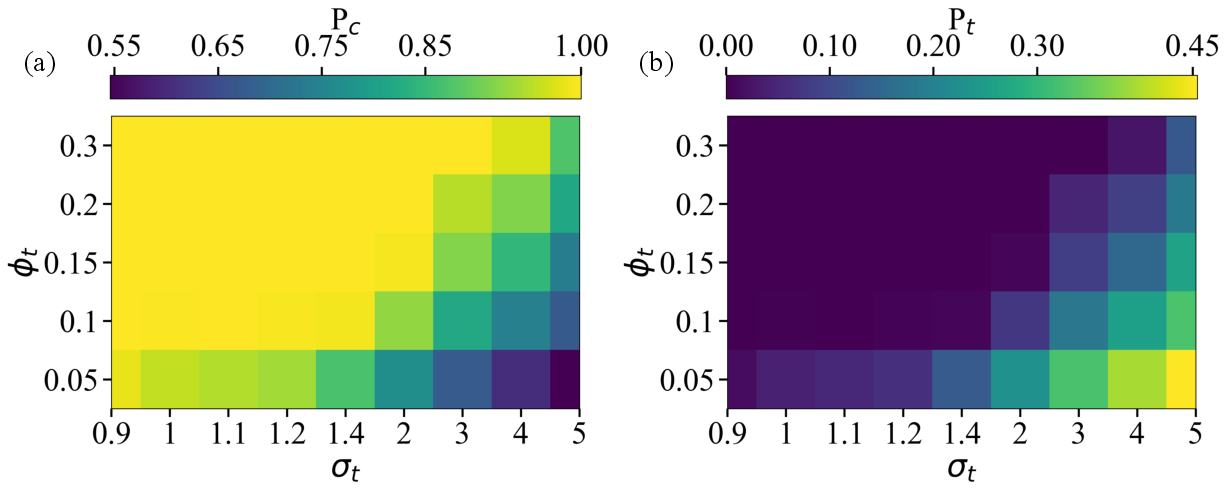}
  \caption{ The plot elucidates how translocation probabilities $P_c$ and $P_t$ vary as the function of the size of crowders ($\sigma_t$) and their packing fraction ($\phi_t$) in one-sided (trans) crowded environment.  (a) Translocation probability $P_c$ to the free-side (cis). (b) Translocation probability $P_t$ to the crowded side (trans). }
  %Translocation probability shows the variation from the case of unbiased crowding to the one-sided crowding from zero to its saturation value, that is, from zero to 1. 
  \label{fig:heat_map.eps}
\end{figure*}

In Fig. \ref{fig:heat_map.eps} we construct phase plots for $P_c$ and $P_t$ in the $\phi_t - \sigma_t$ plane. $P_c$ is maximum for higher $\phi_t$ and relatively smaller crowders (see Fig. \ref{fig:heat_map.eps}a, $\phi_t > 0.1$, $\sigma_t < 1$), indicating translocation towards the free-cis side occurs more readily in a densely crowded medium with smaller crowding particles. While for the case of bigger crowders size $\sigma_t$, the effect of crowders to push the polymer to the free-cis side decreases because of the presence of larger voids in the trans side and probability of translocation to the trans side increases $P_t > 0.$  Fig. \ref{fig:heat_map.eps}b is the complimentary phase plot of Fig. \ref{fig:heat_map.eps}a.

\subsubsection{Translocation time as a function of $\phi$, $\sigma_t$}
Now, we study the effect of packing fraction and crowder size on translocation time. Here, the system of interest consists of both polymer and crowder on the {\it trans}-side. Both entities owe different time scales. In 2D, for a crowder moving with a diffusion constant $D_0= \frac{K_B T}{\xi}$, the time scale associated with the movement to a distance of the order of their size $\sigma$ is 
\begin{equation}\label{eq:15}
    \tau_0= \frac{\sigma^2}{4D_0}
    =\frac{\sigma^2 \xi}{4k_BT},
\end{equation}
where $k_B$ is the Boltzamnn constant and $T$ is absolute temperature. For the bigger crowder of diameter $\sigma_t$, time is $\tau$
\begin{equation}\label{eq:16}
    \tau= \frac{\sigma_t^2}{4D}.
\end{equation}
Therefore, 
\begin{equation}\label{eq:17}
    \tau= \tau_0 \frac{\sigma_t^2}{\sigma_0^2}.
\end{equation}
It shows that the time scale for diffusive motion is proportional to the damping constant and the size of the crowder. For same $\xi$, the $\tau$ increases rapidly on increasing $\sigma_t$ \cite{chen2013dynamics}.

As per the Rouse model, the polymer is represented as a chain having N beads. The diffusion coefficient of the Rouse chain is obtained by Einstein's relation $D_R= \frac{k_B T}{\xi_R}$. The polymer diffuses a distance of the order of its size  during a characteristic time, called Rouse time $\tau_R$ is \cite{doi1988theory,rubinstein2003polymer,chen2013dynamics}
\begin{equation}\label{eq:18}
    \tau_R= \tau_0 N^{(1+2\nu)},
\end{equation}
where $\nu$ is the flory exponent. As polymers exhibit self-similarity, they are characterized by $N$ distinct relaxation modes. Each relaxation mode is designated by a mode index $p=1, 2, 3, \cdots, N$  can be written as 
\begin{equation}\label{eq:19}
    \tau_p= \tau_0 \frac{N}{p}N^{(1+2\nu)}.
\end{equation}
Within the system, the polymer is a large object that moves slowly, while the crowder movement depends on their size. In cases where the time scales for the crowder is shorter than $\tau_p = \tau_0$ for $p=N$, the correlation between the crowders is insignificant, and they provide a uniform random background for the polymer. However, when the two-time scales are comparable, as in this study, the polymer's motion correlates with the crowder's. As the size of the crowder increases, their correlations become stronger. On increasing the packing fraction of crowders, the diffusion coefficient of polymer and crowder decreases, which means correlations between the crowders become stronger.

\begin{figure*}
  \includegraphics[width=\textwidth]{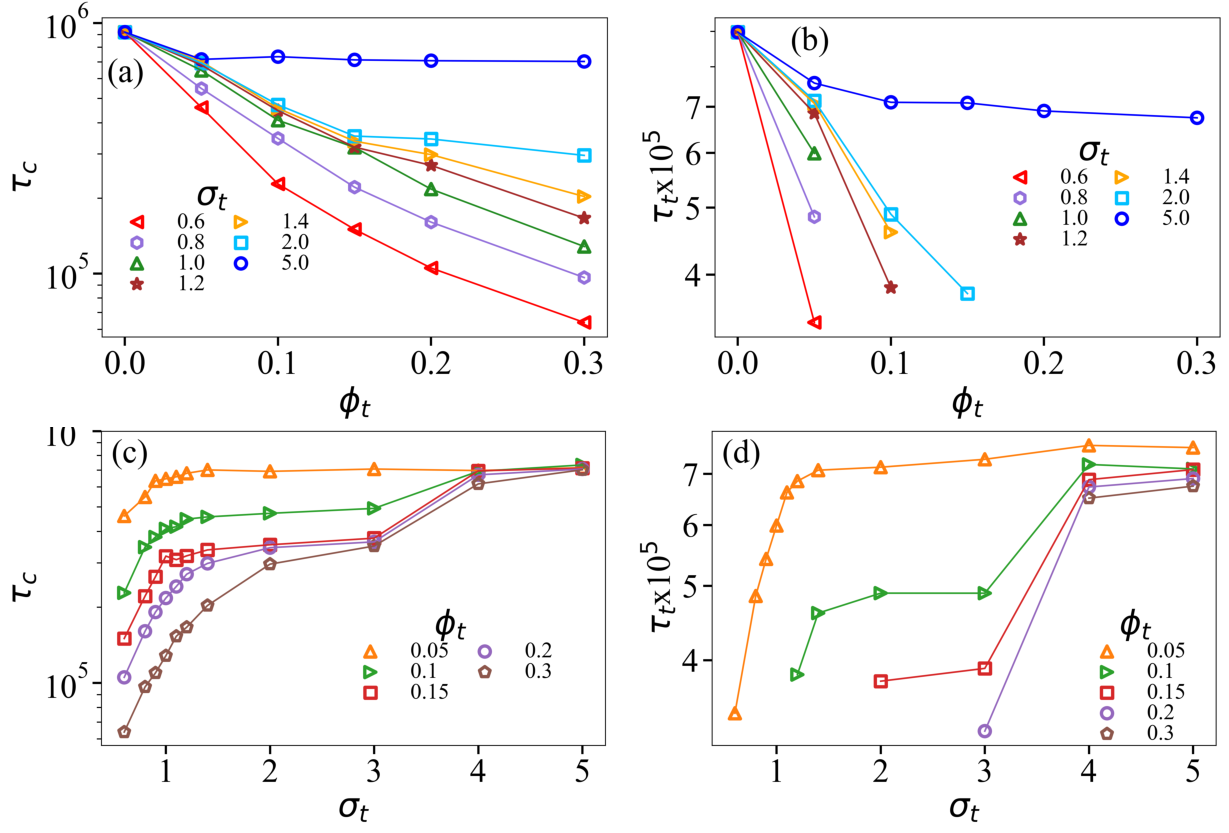}
  \caption{Upper Panel: Translocation time as a function of packing fraction of inert crowders $\phi_t$ on the trans-side. (a) Translocation time $\tau_c$ to the free cis-side for different $\sigma_t$. (b) Translocation time $\tau_t$ to the crowded trans-side for different $\sigma_t$.
  Lower Panel: Translocation time as a function of size of inert crowders $\sigma_t$ on the trans-side. (c) Translocation time $\tau_c$ to the free cis-side for different $\phi_t$. (d) Translocation time $\tau_t$ to the crowded trans-side for different $\phi_t$.}
  \label{fig:time_oneside.eps}
\end{figure*}

$\tau_t$ and $\tau_c$ is the time the polymer takes to translocate to the crowded trans side and to the free cis-side. In Fig. \ref{fig:time_oneside.eps}a and \ref{fig:time_oneside.eps}b we have plotted $\tau_c$ and $\tau_t$ as a function of $\phi_t$ for different $\sigma_t$. $\tau_c$ decreases monotonically with increasing $\phi$. A decrease in overall $\tau_c$ can be understood in terms of mean entropic force. On this basis, an increase in $\phi_t$ and random distribution of crowders on the {\it trans}-side will lead to lower entropy on this side and drive the polymer to translocate to the side of no crowder having higher entropy. Hence, in less time, $\tau_c$ as compared to $\tau_t$ polymer gets translocated to no crowding side. A sudden jump in the graph for higher $\sigma_t$ can be interpreted from Fig. \ref{fig:time_oneside.eps}c, d. It shows that translocation time is a function of $\sigma_t$ for different $\phi_t$. $\tau_c$ increases with an increase in $\sigma_t$ and a decrease in $\phi_t$. The increase in $\tau_t$ is more rapid as compared to the $\tau_c$. For a fixed $\phi_t$, on larger $\sigma_t$, the case of translocation to {\it trans}-side becomes less probable, leading to higher translocation time. Specifically, for higher $\sigma_t$, that is, $\sigma_t \geq 3$, the increase in the size of crowders on the {\it trans}-side leads to a rare, large void and more space for the polymer to move. Hence, the $\tau_t$ shows a sudden increase in its value. Also, in terms of osmotic pressure, as the number of crowders increases, the force exerted by them on the polymer also increases. For a fixed $\phi_t$ and the fixed area of the box on the {\it trans}-side, the $N_t$ has a direct relation with the force exerted by them on the polymer and an inverse relation with the square of the size of the crowder. This also manifests the sudden jump in $\tau_t$ behavior (Fig. \ref{fig:time_oneside.eps}d). Unlike Kaifu and Luo et al. \cite{chen2013dynamics}, there is no resistive force on the polymer translocation (even in the case of one side free crowders where the polymer has more entropy to go to the {\it cis}-side but prefers {\it trans}-side having bigger crowder) in fact polymer translocation to {\it trans}-side is effectively increased for higher $\sigma_t$. 

 %\lipsum[1-2]

\subsubsection{Non zero translocation probability towards the crowders side}

    \begin{figure}[h!]
        \centering
        \includegraphics[width=\linewidth]{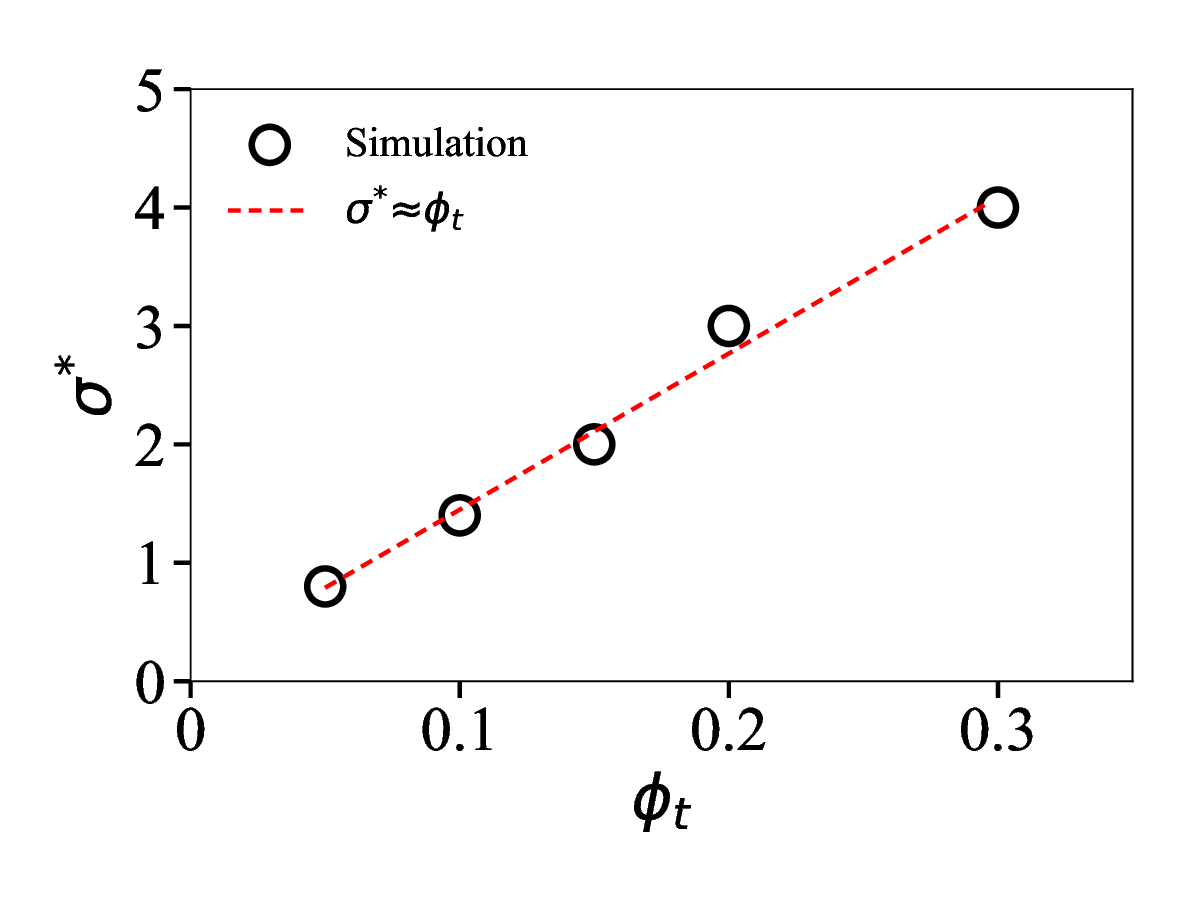}
		%\vspace{20mm}
        \caption{Sigma splitting. Packing fraction of the crowder on trans-side $\phi_t$  as a function of critical value of sigma $\sigma_t \approx \sigma^{*}$, that is, the size of crowders above which polymer have non-zero $P_t$.}
        \label{fig:Sigma_splitting.png}
    \end{figure}

It is expected for one-sided crowders that translocation to the free {\it cis}-side will always be preferred. For small crowding sizes, when the effect of the packing fraction is the strongest, there is almost $100 \%$ certainty that the polymer will translocate to the cis side, leaving $P_t=0$. However, we have already discussed that as we increase the crowding size $\sigma_t$ for the same $\phi$, the effective impact of the crowding environment weakens and gives rise to non-zero $P_t$. For any specific value of $\phi$, there is minimum $\sigma_t$, say $\sigma^{*}$, from where $P_t$ starts to become $P_t > 0$. This minimum value of $\sigma_t$, we define as critical size $\sigma^{*}$ (Fig. \ref{fig:Sigma_splitting.png}). It will be interesting to observe how this critical value of crowder size $\sigma^{*}$ depends on the packing fraction $\phi$.  Note from Fig. \ref{fig:prob_oneside.eps}c that these $\sigma^{*}$, the critical values  of $\sigma_t$,  increase with  $\phi_t$ and follows an almost linear relation, $\phi_t \approx \sigma^{*}$. This can be understood on the basis of entropy where the voids created on crowded {\it trans}-side provide a large room and hence higher entropy for the polymer to explore than from {\it cis}-side, which results in a non-zero $P_t$.

\subsubsection{ Effect of asymmetric initial configuration of polymer on translocation probability}
\begin{figure*}
        \centering
        \includegraphics[width=\textwidth]{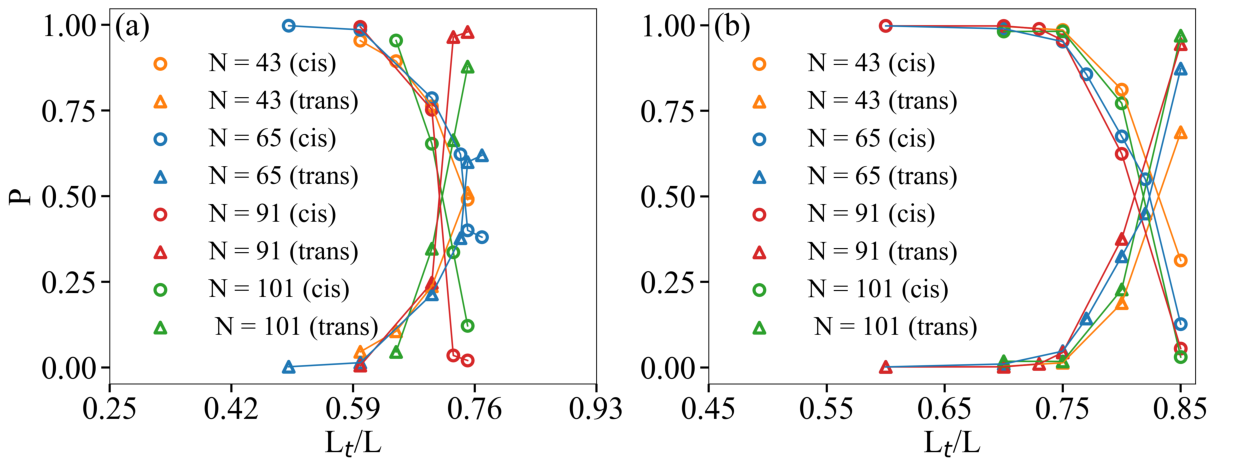}
        \caption{Shifing of polymer length in the crowded trans-side.(a) Sparse crowding $\phi_t = 0.1$ (b) Comparatively dense crowding $\phi_t = 0.2$ with polymer length, \textit{N} = 43, 65, 91, and 101.} %Points where shifted polymer length crosses trans and cis probability are 0.73, 0.745, 0.75.
        \label{fig:shifting_pf_0.1.png}
    \end{figure*}

\begin{figure}[ht!]
        \centering
        \includegraphics[width=\linewidth]{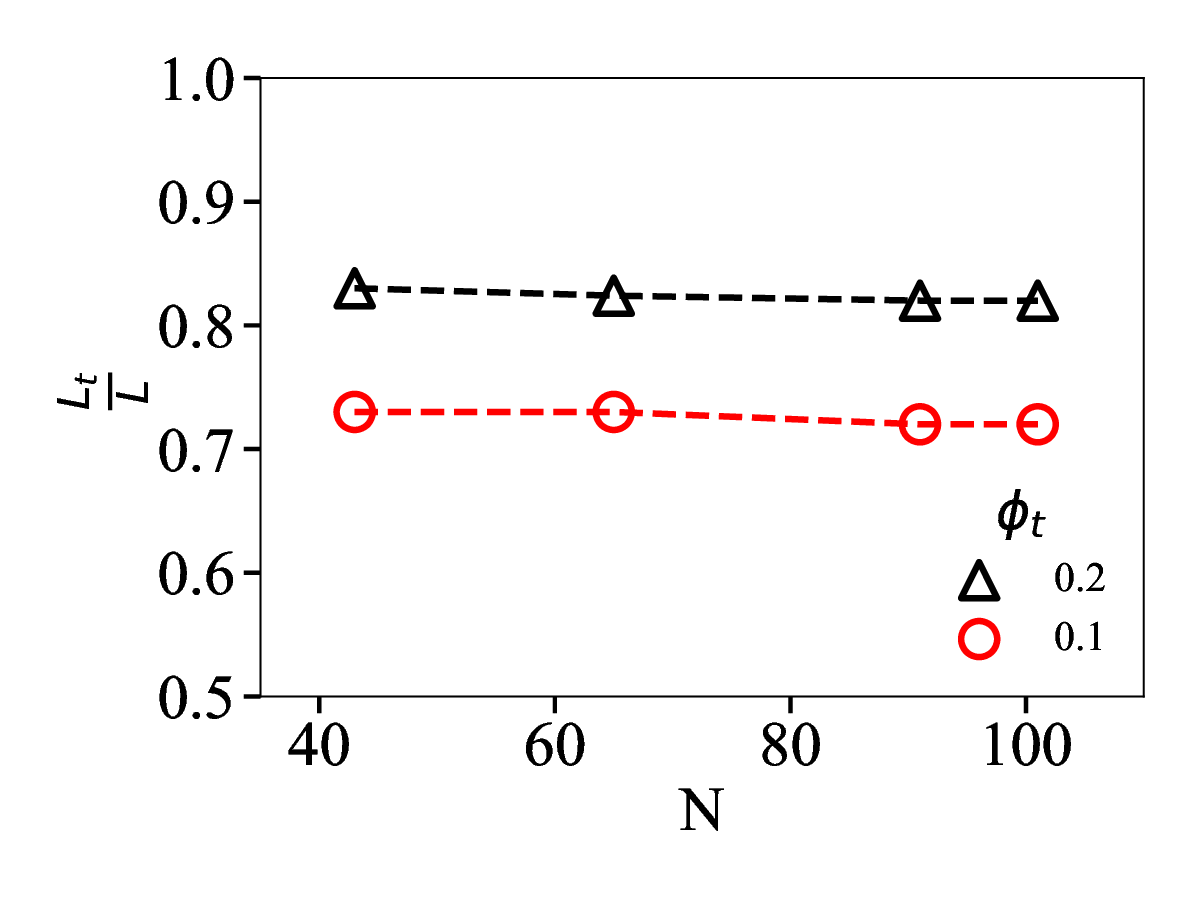}
		%\vspace{20mm}
        \caption{Combined plot of polymer length shifted $L_t/L$ to crowded environments for two different $\phi_t = 0.1, 0.2$ against the original length of the polymer $L$. Both the plots show consistent behavior for a particular value of $\phi_t$ as a function N.}
        \label{fig:Plots/Shifting_all.png}
    \end{figure}

For the initial part of our study, we investigated the translocation dynamics of a polymer through a pore in the presence of free and one-sided crowder. Specifically, we have examined the dynamics of the polymer when it passes with its middle monomer at the pore between {\it cis} and {\it trans} side of the box. In the free case, when there are no crowders, the environment is symmetric, and a symmetrically placed polymer with a middle monomer at the pore exhibits an equal probability of translocation to either side ($P_c = P_t = 0.5$). The introduction of crowders on one side of the box breaks the symmetry in the translocation dynamics of the polymer, resulting in a preference for translocation in the direction of lower osmotic pressure and higher entropy ($P_c = 1, P_t = 0$)\cite{bhattacharjee2013flory, gopinathan2007polymer, roos2014entropic}. Now, we look at how we can overcome the translocation barrier imposed by the crowders by introducing asymmetry to its length. To restore the symmetric behavior observed in the absence of crowders, we positioned the polymer asymmetrically on the side with crowders and provided it with an extra shifted length. Our study suggests that the additional length of the polymer provides an edge to the crowders, allowing them to bias the direction of translocation more effectively.  Fig. \ref{fig:shifting_pf_0.1.png} represents the translocation probability $P$ to either side with the ratio of shifted length $L_t$ of the polymer of length $L$. The first case is a polymer shifting from a sparsely crowded environment of $\phi$= 0.1 (Fig. \ref{fig:shifting_pf_0.1.png}a) with different lengths to a densely packed crowding $\phi$ = 0.2 (Fig. \ref{fig:shifting_pf_0.1.png}b).  As the polymer shifts to the {\it trans}-side where crowders reside, the likelihood of the polymer crossing over to the free {\it cis}-side begins to decrease from its maximum value. This decrease is due to a reduction in the length of the polymer on the {\it cis}-side, resulting in a decrease in available conformational entropy. A channel exists in the vicinity of the pore. The cylindrical nanochannel created by the crowders provides confinement to the polymer, resulting in a lower available area for the polymer to dangle freely. As we increase the asymmetry by shifting the polymer more to the crowder side, the confinement produced by the crowders in the pore's proximity constrains the motion on the free side due to propagation through the backbone of the polymer. Once an interesting situation arises where the polymer on the crowder side has more freedom to move compared to the confined part of the channel on the {\it cis} -side, which gives the polymer the push to move towards the crowder side.  Howsoever, the length of the polymer, this channel will compensate for its effect on either side of the pore.
This can also be interpreted in terms of entropy, the {\it trans} side being crowded and having more monomers due to shifted length by which configurational entropy increases; hence, the driving force for the polymer to move towards the crowders side increases. Consequently, there is an increase in the $P_t$ value. The two curves converge at a shifted length of $L_t/L$, where the probability of translocation to either side is $50\%$, making the polymer unbiased. On shifting polymer length towards the crowders, they create a higher resistance for the polymer as it tries to translocate through the pore. As the crowding level increases for higher $\phi_t$ (Fig. \ref{fig:shifting_pf_0.1.png}b), the polymer experiences more resistance, and the probability of translocating towards either side becomes balanced at higher shifted length. A comparative plot for both packing fractions comprising of shifted length against the total length of the polymer is shown in Fig. \ref{fig:Plots/Shifting_all.png}.

On the crowded trans side, the free energy equation remains F=E-TS but with a substantial positive E value (repulsive interactions with crowders and polymer). This results in higher free energy compared to the cis side, leading to lower entropy on the trans side. Introducing asymmetry by shifting the polymer to the trans side increases the number of monomers on the trans side, elevating the interaction energy contribution for the shifted length. Simultaneously, the shifted polymer enhances the configurational entropy on the trans side compared to the dangling tail on the cis side. The net effect is a decrease in the free energy of the chain and higher entropy on the trans side, facilitating translocation toward the crowded trans side.

    \begin{figure*}
        \centering
        \includegraphics[width=\linewidth]{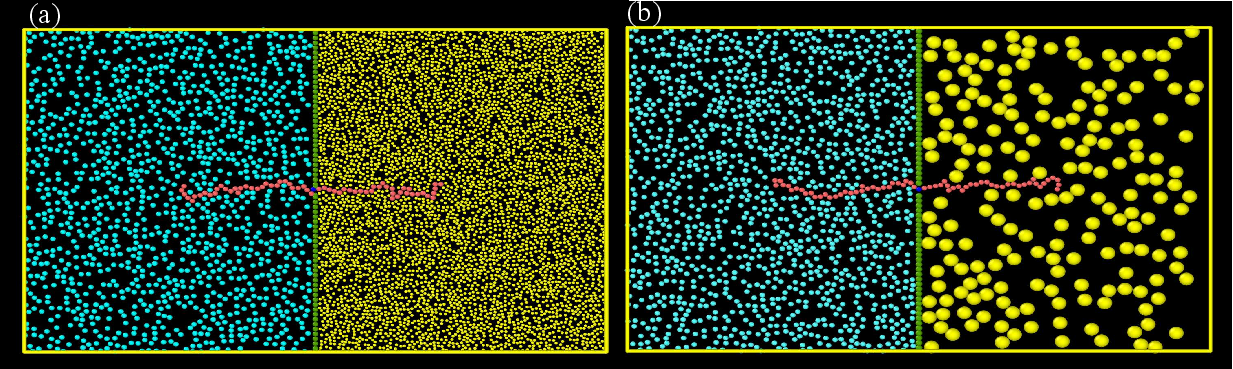}
        \caption{Schematic illustration of polymer translocation process through a pore with asymmetric crowders on both sides of the box. The size of crowders on the cis-side is kept fixed $\sigma_c =1$. The size of crowders on trans-side $\sigma_t$ keeps on changing while keeping the packing fraction on both sides alike ($\phi_t = \phi_c = \phi$). (a) Packing fraction of crowders on both sides $\phi = 0.3$ (densely crowded) while $\sigma_c = 1$ and $\sigma_t = 0.6$ (case: $\sigma_t < \sigma_c$). (b) Packing fraction of crowders on both sides $\phi = 0.3$ while $\sigma_c$ = 1 and $\sigma_t$ = 4 (case: $\sigma_t > \sigma_c$)}
        \label{fig:both_crowder.eps}
    \end{figure*}

\subsection{\textbf{Effect of both sided crowders}} 
    
Now, we study the translocation dynamics of a flexible polymer driven by both side crowders to mimic environments similar to biological situations and in-vitro setups (Fig. \ref{fig:CombinedSchematic}c). In this study, we kept both side packing fractions equal: $\phi_c = \phi_t$ and varied the relative size of the crowders. For simplicity, we always kept the cis-side crowding size fixed at $\sigma_c=1$ and varied the trans-side crowding size $\sigma_t$ (Fig. \ref{fig:both_crowder.eps}). To start, we positioned the middle monomer ($N/2+1$) of the polymer chain at the pore and allowed the beads in the chain to reach equilibrium conformations through thermal collisions. Once the equilibration is completed, the polymer achieves a random configurational state, which is considered to be the initial confirmation for the translocation process. A successful translocation occurred when the polymer chain ended up on either side of the pore within the simulation time. For successful translocations, we have observed translocation probability and time as a function of packing fraction $\phi$ and crowding size $\sigma_t$. This phenomena of polymer translocation driven by crowders can be elucidated by quantifying the average distance among the crowders $ \langle r_s \rangle $ and mean entropic force  $ \langle F_r \rangle $. 

Consider that there are $N_c$ crowders in general of the diameter $\sigma_i$ on either side of the box with dimension ($L_x \times L_y$), then the packing fraction is 
\begin{equation}\label{eq:20}
  \phi= \frac{N_c ~ \pi (\frac{\sigma_i}{2})^2}{L_x \times L_y}.  
\end{equation}
Using Eq.(17) and alike Eq.(7), the average distance between the surface of the crowders $ \langle r_s \rangle$ is 
\begin{equation} \label{eq:21}
   \langle r_{s} \rangle = \sigma_i \left( \frac{1}{2} \sqrt{\pi/\phi}-1 \right),
\end{equation}
The packing fraction of crowders at both sides is set as  $\phi_t = \phi_c = \phi$, and the size of the crowder at the cis-side is kept fixed $\sigma_i = \sigma_c = 1$. On the trans side, the size of the crowder varies $\sigma_t = \sigma_i$ within $0.6 \leq \sigma_t \leq 2.5$ for our case. Thus, the average distance between the surface of the cis-side crowders $\langle r_{cs} \rangle$ and trans-side crowder $\langle r_{ts} \rangle$ is 
\begin{equation} \label{eq:22}
   \langle r_{cs} \rangle = \sigma_c \left( \frac{1}{2} \sqrt{\pi/\phi}-1\right),  ~~~ \langle r_{ts} \rangle = \sigma_t \left( \frac{1}{2} \sqrt{\pi/\phi}-1\right).
\end{equation}
Eq. \ref{eq:21} and Eq. \ref{eq:22} indicates that for a fixed $\phi$, $ \langle r_s \rangle$ increase with increasing $\sigma_i$ and it decreases with increasing $\phi$ for a fixed $\sigma_i$. It will provide insight into understanding the phenomena of translocation induced by crowders and the average distance between their surfaces.

As the polymer is coming out of a pore, crowders create a confined cylindrical channel of size $ \langle r_s \rangle$. For this mean entropic force  $ \langle F_r \rangle$ can be written as\cite{roos2014entropic, chen2013dynamics, muthukumar1989effects}
\begin{equation} \label{eq:23}
     \langle F_r \rangle =  F_0 \frac{k_BT}{\langle r_s \rangle}, 
 \end{equation}
 where $F_0$ is the proportionality constant, $k_B$ is the Boltzmann constant, $T$ is the absolute temperature.
 
 Simplifying the expression of $ \langle F_r \rangle$ for cis and trans-side crowder in terms of $\langle r_{cs} \rangle$, $\langle r_{ts} \rangle$, $\sigma_c$, $\sigma_t$, and $\phi$ gives
  \begin{equation} \label{eq:24}
     \langle F_r \rangle  =  F_0 k_B T  \left (\frac{1}{ \langle r_{cs} \rangle} - \frac{1}{\langle r_{ts} \rangle}\right ) ,
 \end{equation}
 \begin{equation} \label{eq:25}
    \langle F_r \rangle = F_0 k_BT {\frac{1}{\frac{1}{2} \sqrt{\pi/\phi}-1 }\left (\frac{1}{ \sigma_{c}} - \frac{1}{\sigma_{t}}\right).}
 \end{equation}
Eq. \ref{eq:25} show that $ \langle F_r \rangle$ depends on $\phi$ and varying trans-side crowders of $\sigma_t$. With increasing $\phi$ or changing $\sigma_t$ keeping $\sigma_c$ fixed, $ \langle F_r \rangle$ increases \cite{chen2013dynamics}. Polymer prefers to move the size of bigger crowders due to larger conformational entropy.
 
 Further, Osmotic pressure ($\Pi$), which is a colligative property, depends on the number density of the crowder and influences the translocation dynamics. We put a polymer chain at the center of the wall, separated by a difference in crowding size, which creates a crowding concentration gradient on either side. The polymer chain will be driven by the difference in osmotic pressure created by the concentration gradient and prefers to stay on the side of relatively lower osmotic pressure and higher volume. Thus, the competition between mean entropic force and osmotic pressure determines the direction of the translocation process.
 
\subsubsection{Translocation probability as a function of $\phi$, and $\sigma_t$}

%\textcolor{magenta}{\it{\textbf{NEW:} Switch from $P_t=0$ to $1$ as we cross $\sigma=1$, same with $P_c$. Strongness of the tiny crowders.}}  \newline

\begin{figure*}
  \includegraphics[width=\textwidth]{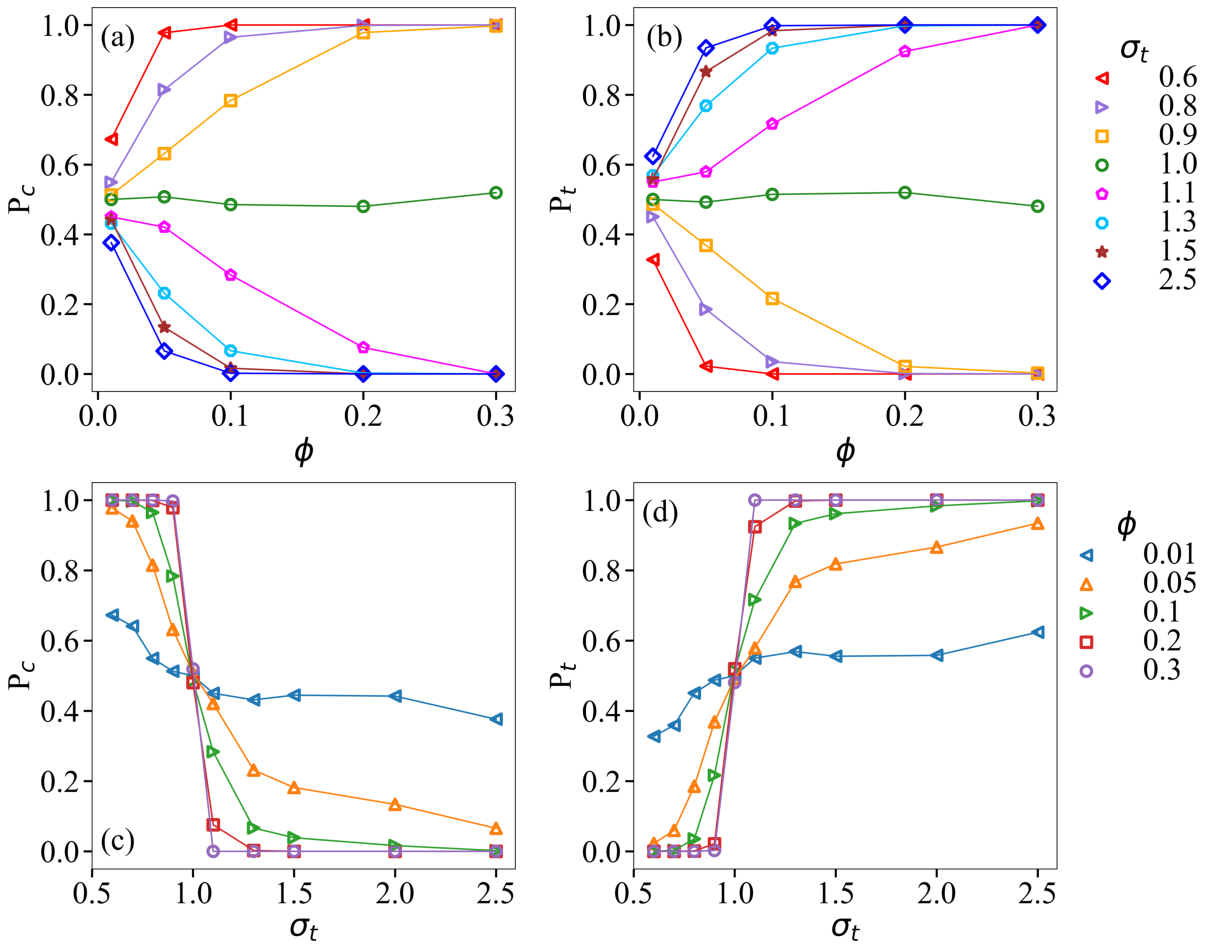}
  \caption{Upper panel: Translocation probability as a function of packing fraction $\phi$. (a) Translocation probability $P_c$ to cis-side for different $\sigma_t$. (b) Translocation probability $P_t$ to trans-side for different $\sigma_t$. Lower panel: Translocation probability as a function of $\sigma_t$ of inert crowders on the trans side. (c) Translocation probability $P_c$ to cis-side for different $\phi_t$. (d)  Translocation probability $P_t$ to trans-side for different $\phi_t$.}
  \label{fig:prob_bothside.eps}
\end{figure*}

\begin{figure*}
  \includegraphics[width=\textwidth]{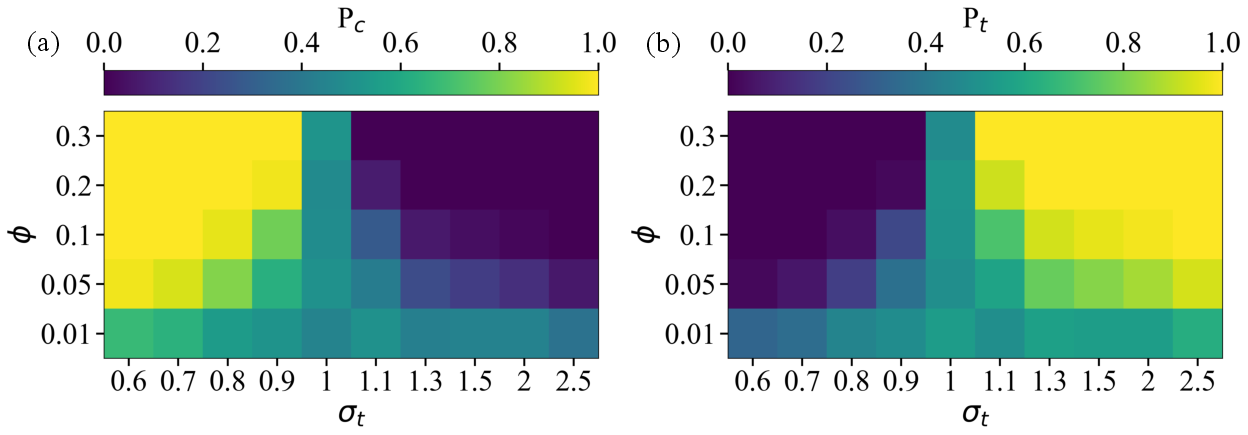}
  \caption{The plot elucidates how translocation probabilities $P_c$ and $P_t$ vary as the function of the size of crowders on the trans side ($\sigma_t$) and their packing fraction ($\phi$) in a both-sided crowded environment. The size of crowders on the cis side is kept fixed at $\sigma_c =1$, and the packing fraction of both sides is kept equal, that is, $\phi_c = \phi_t = \phi$. It shows a switch in the translocation of a polymer by tuning crowding size. (a) and (b) Translocation probability to cis-side and trans-side, respectively.
}
  \label{fig:heat_map2.eps}
\end{figure*}

The probability of polymer translocating to the cis-side and trans-side is represented by $P_c$ and $P_t$, respectively. In Fig \ref{fig:prob_bothside.eps}a and Fig \ref{fig:prob_bothside.eps}b, we have plotted the translocation probability of the polymer as a function of the packing fraction $\phi$ for different $\sigma_t$, note $\sigma_c$ is always kept constant at $\sigma_c=1$. Plot shows bifurcation from equally probable scenario ($P_t = P_c = 0.5$ at  $\sigma_t = 1$) to two opposite extremes ( $\sigma_t \neq \sigma_c$ ). For $\sigma_t \leq \sigma_c$, the polymer has a tendency to move to the cis side where bigger crowder resides. $P_c$ rapidly increases from 50-50 $\%$ at lower $\phi$ and then slowly reaches its maximum value at higher $\phi$. While for the case of $\sigma_t \geq \sigma_c$, the polymer has a tendency to move to the trans side. This indicates that for the same $\phi$ on both sides, translocation is preferred to the side of bigger crowders. This behavior is reminiscent of the effect of osmotic pressure, which provides a push to the polymer to translocate from a higher crowder concentration to a lower concentration side. For the same $\phi$, osmotic pressure ($\Pi$) at the side of the smaller crowder is higher than that of the side of a bigger one, leading to the polymer moving to the bigger crowder side. Such behavior of probability with $\phi$ for polymer translocation through pore have been reported earlier \cite{chen2013dynamics}.

We next look at the translocation probability of the polymer as a function of $\sigma_t$ for different $\phi$. $P_t$ displays distinct features as $\sigma_t$ is varied (Fig. \ref{fig:prob_bothside.eps}c, d). For $\sigma_t < \sigma_c$ till it reaches $\sigma_t = 1$, that is, for the relatively small trans-side crowders, polymer translocates to the cis-side ($P_t = 0$). But, the moment there is a crossover of $\sigma_t = 1$, the curve shows a sudden jump in its behavior from its lowest to saturation ($P_t = 1$). This sudden switch prevails over the higher $\sigma_t$ for the highest $\phi = 0.3$ and shows a fall from $P_t = 1$ for decreasing $\phi$. Additionally, even for $\phi = 0.01$, which is almost a crowder-free environment, the strong effect of tiny crowders ($\sigma_t < \sigma_c$) pushes the polymer to translocate to the cis-side ($P_t = 0.3$). While for the same $\phi = 0.01$ at $\sigma_t > \sigma_c$, the polymer has an equal probability of moving to either side as it is like a tug-of-war situation. This result can be interpreted in terms of the mean radial entropic force. $\langle F_r \rangle$ increases with increasing $\sigma_t$ and $\phi$ and hence leads the translocation dynamics. Unlike Chen et al., 2013 \cite{chen2013dynamics}, our model has no bottleneck or resistive force pertaining to non-monotonic behavior.\\
The results in Fig. \ref{fig:prob_bothside.eps} are relevant to the simulation study by Chen and Luo\cite{chen2013dynamics}. They observed that the polymer with an initial configuration where the middle monomer at the pore on the wall separating two sides will prefer to translocate to the side of bigger crowders. They note that polymer that goes to the side of a bigger crowder exhibits a maximum probability on increasing $\sigma$ up to a certain extent and then shows a downfall due to the interplay of the driving force and the resistive force, specifically from the bottleneck leading to the non-monotonic behavior of probability. They observed that at large $\sigma$, it shows the resistive force of large magnitude compensated by entropic force leads to a decrease in probability to the bigger crowder side. While, in our case, the calculations are consistent with Chen and Luo and likewise to Polson \cite{polson2019polymer}, our model also contradicts this and denies any bottleneck in the system pertaining to non-monotonicity in the curve. In our case, the probability curve exhibits a continuous increase as the polymer prefers to translocate to the side of bigger crowders and does not account for the downfall in the curve is noted.

The behavior of the crossover region of $\sigma_t = 1$ as a function of $\phi$ is presented in phase plot for $P_c$ and $P_t$ in $\phi - \sigma_t$ plane (Fig. \ref{fig:heat_map2.eps}). $P_c$ is maximum for smaller $\sigma_t$ and higher $\phi_t$ (see Fig. \ref{fig:heat_map2.eps}a, $\sigma_t < 1, \phi > 0.05$). For relatively bigger crowders, the phase plot shows a uniform increase in $P_t$ with increasing $\sigma_t$ for all $\phi$ (see Fig. \ref{fig:heat_map2.eps}b).

\begin{figure*}
  \includegraphics[width=\textwidth]{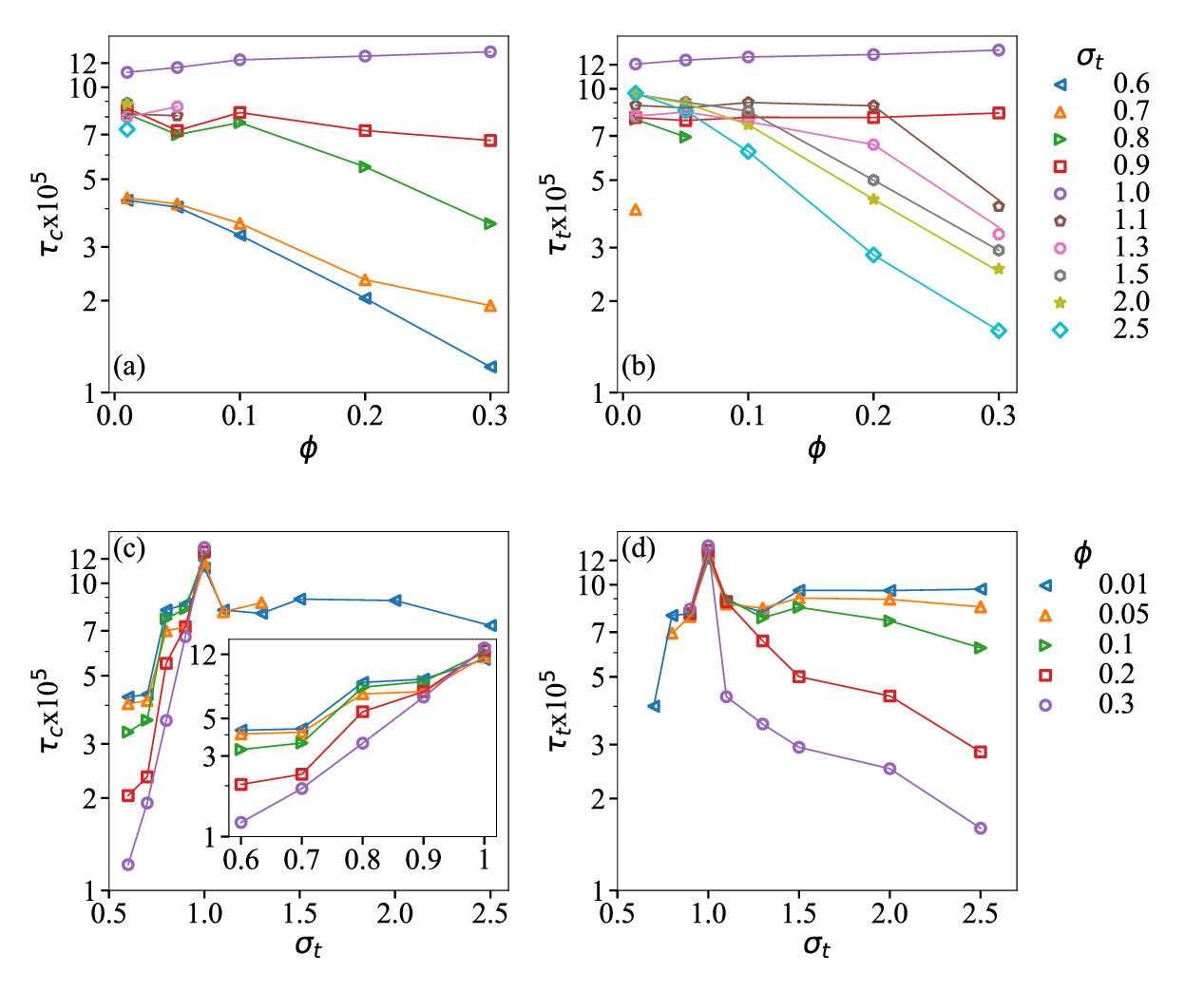}
  \caption{ Upper Panel: Translocation time as a function of packing fraction $\phi_t$. (a) Translocation time $\tau_c$ to the cis-side for different $\sigma_t$. (b) Translocation time $\tau_t$ to the trans-side for different $\sigma_t$.
  Lower Panel: Translocation time as a function of size of inert crowders $\sigma_t$ on the trans-side. (c) Translocation time $\tau_c$ to the cis-side for different $\phi_t$. (d) Translocation time $\tau_t$ to the trans-side for different $\phi_t$.}
  \label{fig:Translocation_time_phi.eps}
\end{figure*}

\subsubsection{Translocation time as a function of $\phi$, and $\sigma_t$}

We first look at the translocation time as a function of packing fraction $\phi$ for different values of crowder size $\sigma_t$ (Fig. \ref{fig:Translocation_time_phi.eps}a, b). Owing to an alike environment on both sides of the pore for $\sigma_t = \sigma_c = 1$, the overall $\tau$ owes the highest value. Further diverging from $\sigma_c = \sigma_t =1$ and moving to the asymmetric nature of crowded size ($\sigma_c \neq \sigma_t$), we observe a decrease $\tau_c$ and $\tau_t$ with increasing $\phi$ for both the situation: $\sigma_t < \sigma_c$, and $\sigma_t > \sigma_c$. This result can be elucidated in the context of $\langle F_r \rangle$, which increases on increasing $\phi$. For $\sigma_t < \sigma_c$, the combined effect of osmotic pressure $\Pi$ and mean entropic force $ \langle F_r \rangle$ moves the polymer to lower crowding concentration and higher entropy which dominate the translocation phenomenology and drives the polymer on bigger crowder side of the pore ensuing lower $\tau_c$  (Fig. \ref{fig:Translocation_time_phi.eps}a) and for $\sigma_t>\sigma_c$ holds the higher value of the average distance between the surface of crowder on trans-side than on the cis side, providing more space for the polymer to translocate easily in lesser time $\tau_t$ (Fig. \ref{fig:Translocation_time_phi.eps}b).
Additionally, translocation is a stochastic process, and random fluctuations could lead the polymer to move to either the cis or trans side. But it's not necessary that translocation would occur for all particular cases, and pointed out translocation at these points is a very rare event since we introduced biases in the form of crowders in extreme bias, we wouldn't observe any translocation event taking place to the unfavoured side. In Fig. 19a, translocation time values corresponding to $\sigma_t = 0.8, 2.0$ are not present at certain values of $\phi$. This absence is attributed to the rare and stochastic nature of translocation events. The lack of translocation time values on either side indicates the non-occurrence of translocation in that specific direction. To elaborate, when translocation doesn't occur on the cis side at a given time, but the polymer successfully translocates to the trans side, the corresponding translocation time values are reflected in Fig. 19b.
We next look at the situation for translocation time varying with $\sigma_t$ for different $\phi$ (Fig. \ref{fig:Translocation_time_phi.eps}c, d). At lower $\sigma_t < \sigma_c$, $\tau_c$ rises faster for higher $\phi$. On the other hand for $\sigma_t > \sigma_c$, $\tau_t$ shows decreasing behaviour on increasing $\phi$. As explained earlier, the combined effect of $\Pi$, $\langle F_r \rangle$, $\langle r_s \rangle$ pulls the polymer rapidly to the side of bigger crowder (Fig. \ref{fig:Translocation_time_phi.eps}c and Fig. \ref{fig:Translocation_time_phi.eps}d). 

\begin{figure*}[ht]
        \centering
        \includegraphics[width=\linewidth]{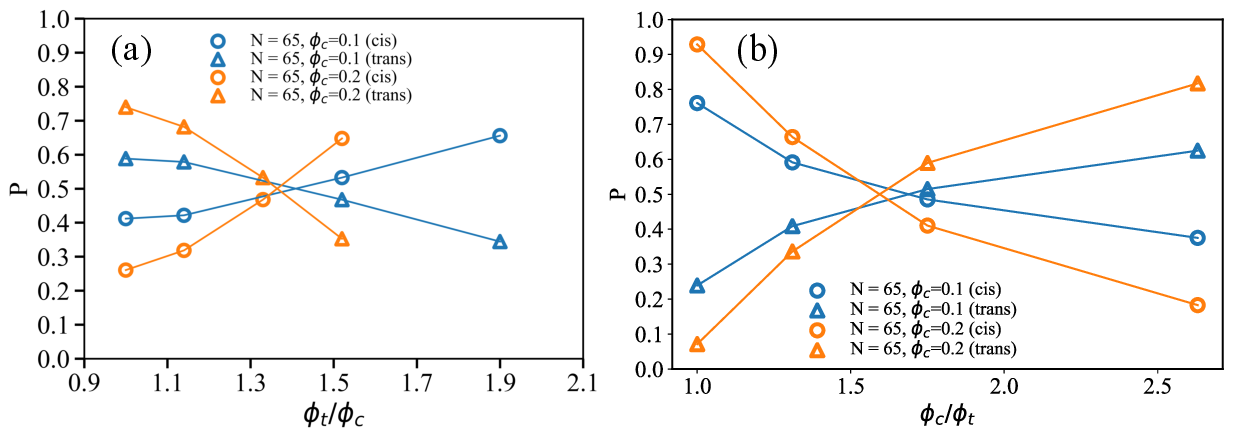}
        \caption{(a)Translocation probability $P$ as a function ratio of packing fraction of crowders on the trans-side and cis side $\phi_t/\phi_c$ for two distinct values of $\phi= 0.1, 0.2$. The size of crowder on the cis side $\sigma_c = 3$ while that on the trans side $\sigma_t = 4$. (b) The size of crowder on the cis side $\sigma_c = 3$ while that on the trans side $\sigma_t = 2$. }
        \label{fig:all_in_one_n_terms_combined.png}
    \end{figure*}

%\subsubsection{\textcolor{red}{NEW}:\textbf{ Get back 50-50 by changing the number of crowder for a fixed $\phi$.}}

\subsubsection{Effect of crowding size on translocation probability}

Translocation for the case of no crowders provides us the $50-50\%$ probability ($P_c \approx P_t \approx 0.5$). Initially, the symmetry of the environment of obtaining equal probable case has been broken by introducing crowders on one side ($P_c = 1, ~P_t = 0$). We have observed that most of the time polymer is translocating to the side of the crowders on the grounds of mean entropic force, osmotic pressure, and distance between the surface of the crowders. Next, the symmetry has been broken by introducing asymmetry in the initial configuration of the polymer, that is, shifting the length towards the crowder side. A ratio of $L_t/L$ exists where the equal probable case of cis and trans-translocation is retrieved.

Now, when the crowders are being introduced on both sides of the box, we have investigated that the polymer is translocating almost to the side of bigger crowders ($P_c = 0, P_t = 1$ for $\sigma_t < \sigma_c$, and $P_c = 1, P_t = 0$ for $\sigma_t > \sigma_c$). Similarly, in this case, to restore an equal probability scenario, symmetry is broken by changing the packing fraction of crowders on the trans side, i.e., $ \phi_c$ is not equal to $\phi_t$ anymore. We are choosing a particular value of the $\sigma_t$ and varying the number of crowders $N_t$, which in turn changes the packing fraction $\phi_t$. The goal is to obtain an optimum value of the packing fraction of crowders on the trans-side so that the polymer translocates equally to both sides rather than having a preference for the side of the bigger crowder.  Fig. \ref{fig:all_in_one_n_terms_combined.png}a and Fig. \ref{fig:all_in_one_n_terms_combined.png}b illustrate the relationship between the probability of translocation P and the ratio of packing fractions of two sides $\phi_t/\phi_c$. In Fig. \ref{fig:all_in_one_n_terms_combined.png}a, the crowders on the trans-side are larger ($\sigma_c = 4$) than those on the cis-side ($\sigma_c = 3$), while in Fig. \ref{fig:all_in_one_n_terms_combined.png}b, the trans-side has smaller crowders ($\sigma_c = 2$) than the cis-side ($\sigma_c = 3$). Both figures examine sparse and densely crowded environments of packing fraction on the cis-side ($\phi_c = 0.1, 0.2$), with ($\sigma_c = 3$) held constant while $N_c$ is fixed. $N_t$ varies until the polymer becomes unbiased of size, packing fraction, and the number of crowders, allowing it to translocate to either side with equal probability.

\section{Conclusion}

Medium-driven controlled polymer translocations without any explicit external forces are fundamental problems in science and engineering. Using Langevin dynamics simulations, we extensively study the effect of crowding and polymer length distribution on the translocation processes in crowd-free and crowded environments. We investigated the scaling properties of translocation probability $P$, time $\tau$, and MSD. First, we compared the scaling properties with the results generated by the standard scaling theories of polymer physics. We obtained scaling for translocation time $\tau \sim N^{2.5}$ and when $\tau > \tau_R$ MSD shows $\langle \Delta r^{2} (t)\rangle \sim t^{0.8}$, in free environment. Our results exactly match the experimental data and show subdiffusive behavior. A simple analysis of the average crossing time $\langle \tau_{crossing}\rangle$ of individual monomer, as it crosses the pore, has been performed and laid the foundation of two interesting but counterintuitive features of translocation rate $k_T$ and bead velocity $v_b$. It has been observed that the translocation rate is minimum when the middle monomer passes the pore and keeps on increasing, whereas it holds the opposite behavior for bead velocity, which is maximum for the middle monomer and decreases for the subsequent part of tail monomers. Further, we rationalize the translocation phenomena in the light of nucleation processes. An asymmetrically placed polymer at the pore towards the receiver side once crosses a significant number of monomers across the pore; that is, when the nucleation barrier has been overcome and free energy maxima is reached, the probability of the polymer going back to the donor side is almost negligible.

For one-sided crowding, with crowding on the trans side and free on the cis side, the translocation probability and time have been noted. $P_t$ shows a monotonic decrease with increasing $\phi$ but shows an increase when the crowder's size  $\sigma_t$ is large enough to create rare, large voids the polymer can explore. This provides us with a critical value of $\sigma_t$ called $\sigma^{*}$ where the polymer has non-zero $P_t$ and translocates to the crowded side. A complimentary behavior is also followed by $P_c$.  A breakthrough in the existing model has been achieved by simply shifting polymer length from the middle monomer to the crowded side to examine whether there can be a situation if a crowd-free environment probability can be retraced. The study has been done for both sparse environment $\phi_t = 0.1$ and relatively dense $\phi_t = 0.2$ and we investigated that there exists a ratio of length shift to the polymer length $L_t/L$ where polymer exhibits $P_t \approx P_c \approx 0.5$. Translocation time $\tau_t$ decreases rapidly with increasing $\sigma_t$ and an increase with larger $\sigma_t$.

In the case of two-sided crowding, where the packing fraction of both sides is set fixed, we kept the crowding side on the cis side $\sigma_c$ fixed, and on the trans side, $\sigma_t$ is varying. The translocation probability shows a bifurcation from the equal probable case, and polymer translocates to the side of bigger crowders, $P_t> 0.5$ for $\sigma_t > \sigma_c$  and $P_t < 0.5$ for $\sigma_t < \sigma_c$. The probability curve shows a sudden switch from its lowest to peak value when a crossover of $\sigma_t = \sigma_c$ happens.  Translocation time $\tau_t$ decreases with increasing $\phi$ and $\sigma_t$, indicating the polymer preference to stay towards bigger crowders in less time. A study on changing the relative packing fraction of crowders on both sides leads to the $50-50$ probability akin to the crowd-free situation is done. These results can be interpreted by the interplay between mean entropic force and effective osmotic pressure of the crowders.

   In the case of driven translocation, when an external force is applied on the polymer up to its certain length and after that, if the force is switched off, then the entropy-driven force and free energy are sufficient to keep the polymer on the preferred side. Our model is generally enough to switch the translocation direction without any explicit force. Once the nucleation barrier is crossed for successful translocation, the polymer has an almost extremely low probability of going back to the donor side. In other words, the regulation of the free energy barrier by tuning the polymer length can have significance in the translocation phenomena having biomedical applications, for example, in the field of controlled drug delivery systems. Further, both spatial and temporal control over the translocation of a polymer tethered to a surface can be incorporated into our model. In the context of targeted drug delivery systems, we aim to deliver drugs specifically to designated cells, organelles, etc. This is the direct and potential application of polymer translocation, allowing us to control the direction of polymer translocation in accordance with the target's environment \cite{ bai2020nanoparticle, zhang2012uptake, maitra2023acidic}.
   Controlled translocation of single molecules can be achieved by maintaining control over the nanopore in such a way that we have a speed tuning of individual beads and molecules as well as passing through it at the desired time; that is, molecule-independent speed control can be achieved \cite{leitao2023spatially}.

\begin{acknowledgments}
We would like to thank Jaeoh Shin, Praveen Bommineni, Kunal Rai, and Thyageshwar Chandran for useful discussion and critical comments. VG and SKG would like to thank SERB, DST (Project No. SRG/2020/001606 ) for funding. SKG would like to thank SERB, DST (Project No. MTR/2022/000833) for financial support. We would like to acknowledge NIT Warangal for the research seed funds and institute fellowship to VG. 
\end{acknowledgments}

%=========================================================================
%=========================================================================

\bibliography{main_text} 
\end{document}